\documentclass[prb,twocolumn,superscriptaddress,showpacs,eqsecnum]{revtex4-1}

\usepackage[utf8x]{inputenc}
\usepackage[pdftex]{graphicx}
\usepackage{amsmath,amssymb}
\usepackage{subfigure}

\newcommand{\sgn}{\text{sgn}}

\begin{document}

\title{Weak localization and magnetoresistance in a two-leg ladder model}

\author{Michael P. Schneider} \affiliation{Institut f\"ur Theorie der
  Kondensierten Materie and DFG Center for Functional Nanostructures,
  Karlsruher Institut f\"ur Technologie, 76128 Karlsruhe, Germany}
  \affiliation{Institut f\"ur Nanotechnologie, Karlsruher Institut f\"ur Technologie, 76021 Karlsruhe, Germany}
  \affiliation{Max-Born-Institut, Max-Born-Str. 2A, 12489 Berlin, Germany}
\author{Sam T. Carr} \affiliation{Institut f\"ur Theorie der
  Kondensierten Materie and DFG Center for Functional Nanostructures,
  Karlsruher Institut f\"ur Technologie, 76128 Karlsruhe, Germany}
\author{Igor V. Gornyi} \affiliation{Institut f\"ur Theorie der
  Kondensierten Materie and DFG Center for Functional Nanostructures,
  Karlsruher Institut f\"ur Technologie, 76128 Karlsruhe, Germany}
  \affiliation{Institut f\"ur Nanotechnologie, Karlsruher Institut f\"ur Technologie, 76021 Karlsruhe, Germany}
  \affiliation{A.~F.~Ioffe Physico-Technical Institute, 194021, St.~Petersburg, Russia}
\author{Alexander D. Mirlin} \affiliation{Institut f\"ur Theorie der
  Kondensierten Materie and DFG Center for Functional Nanostructures,
  Karlsruher Institut f\"ur Technologie, 76128 Karlsruhe, Germany}
  \affiliation{Institut f\"ur Nanotechnologie, Karlsruher Institut f\"ur Technologie, 76021 Karlsruhe, Germany}
  \affiliation{Petersburg Nuclear Physics Institute, 188300 St.~Petersburg, Russia}

\begin{abstract}
We analyze the weak localization correction to the conductivity of a spinless two-leg ladder model in the limit of strong dephasing $\tau_\phi \ll \tau_\mathrm{tr}$, paying particular attention to the presence of a magnetic field, which leads to an unconventional magnetoresistance behavior.  We find that the magnetic field leads to three different effects: (i) negative magnetoresistance due to the regular weak localization correction (ii) effective decoupling of the two chains, leading to positive magnetoresistance and (iii) oscillations in the magnetoresistance originating from the nature of the low-energy collective excitations.  All three effects can be observed depending on the parameter range, but it turns out that large magnetic fields always decouple the chains and thus lead to the curious effect of magnetic field enhanced localization.
\end{abstract}


\date{\today}

\maketitle

\section{Introduction}
\label{sec:introduction}

The interplay of interactions and disorder in electronic systems confined to move in one spatial dimension has remained a fascinating theoretical problem for many decades.  It has long been known that any electron-electron interactions in a clean one-dimensional wire destroy the Fermi-liquid which is paradigmatic of higher dimensions and transform the system to a Luttinger liquid which has no coherent fermionic modes.\cite{LL,Haldane-1981}  On the other hand, an arbitrarily small disorder potential in the non-interacting system leads to the localization of all states \cite{Mott-Twose-1961,Berezinskii-1973,Abrikosov-Ryzhkin-1978} leading to strictly zero conductivity at any temperature.  Since these pioneering works, a lot of progress has been made in understanding the effects of both disorder and interactions in one-dimensional systems (in the following, we shall restrict the discussion to repulsive interactions).  Early works\cite{Apel-Rice-1982,Giamarchi-Schulz-1988} focused on the enhancement of the backscattering off impurities induced by the Luttinger liquid, an effect seen already with a single impurity,\cite{Kane-Fisher-1992,Matveev-Yue-Glazman-1993} while recently technical advances have allowed the study of strong localization in the presence of interactions\cite{Gornyi-Mirlin-Polyakov-2005,Basko-Aleiner-Altshuler-2006} -- these works concentrate on quasi-one-dimensional structures, but the same ideas should apply to true one-dimensional systems.  It has even been possible to take the more traditional ideas of weak localization and electron dephasing\cite{Altshuler-Aronov-1985} and apply them successfully to the peculiarities of the one-dimensional geometry.\cite{Gornyi-Mirlin-Polyakov-2007,Yashenkin-Gornyi-Mirlin-Polyakov-2008}

One of the nicest experimental manifestations of weak localization is in the magnetoresistance; in the traditional setting the applied magnetic flux breaks the time-reversal symmetry and thus destroys the weak localization correction to the resistance.  A weak magnetic field meanwhile has relatively little effect on the Drude (diffusive) component on the resistance, and thus allows the quantum weak localization correction to be isolated and studied experimentally.  This unfortunately can not be seen in strictly one-dimensional nanowires however, as in this case the magnetic-field may be gauged out completely and will have no effect (the Zeeman coupling between the magnetic field and the spin of the electrons produces a rather different effect\cite{hurry-up-Andrey}).  However, if one turns from single-chain nanowires to double-chain nanostructures, the so-called ladder models, then interesting orbital effects of the magnetic field may be restored.

While the two-leg ladder is often introduced theoretically as a beautiful model intermediate between one- and two-dimensional systems, it is also rather ubiquitous in nature.  For example, carbon nanotubes which experimentally show power-law scaling behaviour in conductance typical of the Luttinger liquid\cite{nanotube_experiment} are theoretically described in the low-energy regime by a model equivalent to that of a two-leg ladder.\cite{nanotube_theory}  Curious magnetoresistance effects have been seen in nanotubes\cite{nanotube_magnetoresistance} -- however here one should be careful, as the not-completely-trivial mapping from the original nanotube to the two-leg ladder means that the external magnetic flux will couple to the low-energy theory in a more complicated way than for the simple planar two-leg ladder.  There are more direct experimental manifestations of the two-leg ladder however -- including semiconducting double nanowires,\cite{synthetic-double-nanowire} as well as materials which are structurally made up of such ladders, such as PrBa$_2$Cu$_4$O$_8$ where control of disorder and measurement of magnetoresistance are made in experiments.\cite{PrBaCuO}  One can even artificially create such structures using cold atoms in an optical lattice,\cite{Cold_atoms_simulations} the disorder being added as a speckle pattern,\cite{Cold_atoms_disorder} and the effect of a magnetic field being seen on the neutral atoms via the effect of an artificial gauge field.\cite{Cold_atoms_gauge_fields}

Quite generically, ladder models have relevant backscattering terms ($g_1$ terms in the parlance of Luttinger liquid physics), meaning that even the clean system has non-trivial correlations.\cite{Nersesyan-Luther-Kusmartsev-1993,Ledermann-LeHur-2000}  At low temperatures, this leads to a rather surprising response to even a single impurity,\cite{Carr-Narozhny-Nersesyan-2011} and many different possibilities in the presence of a disorder potential.\cite{Orignac-Giamarchi-1997,Orignac-Giamarchi-1996,Orignac-Suzumura-2001}  There has also been a lot of interest in magnetic-field induced phase transitions in the clean ladder systems at low temperature.\cite{Narozhny-Carr-Nersesyan-2005,Carr-Narozhny-Nersesyan-2006,Roux-Orignac-White-Poliblanc-2007,Jaefari-Fradkin-2012}  While these works set the scene for the present study, we will be interested in quite a different regime where we have temperature sufficiently high that we can a) neglect the gaps opened by the interaction backscattering $g_1$ terms; and b) consider sufficiently strong dephasing that the disorder potential may be treated peturbatively.  We will show that even under these conditions, the interplay between the interactions, disorder, and low dimensions gives us curious and counter-intuitive results, such as non-monotonic magnetoresistance, and magnetic-field enhanced localization.  

The paper is organized as follows: in Section II, we will introduce the ladder model that we will analyze.  In Section III we will then study the effect of weak localization and magnetoresistance in two ways: one via a phenomenological introduction of a dephasing time, and the second from a true interacting microscopic model.  We compare the two methods, and show that the phenomenological approach captures most of the essential physics.   The exception is oscillations in the weak localization correction, which have their origin in correlation effects in the ladder model that are equivalent to spin-charge separation.  In Section IV, having calculated the magnetoresistance, we analyze and explain its properties.  Some technical details are relegated to the appendices.  Throughout the paper, we use units where $\hbar=1$.

\section{The ladder model}
\label{sec:laddermodel}

\begin{figure}
 \begin{center}
  \includegraphics[width=0.4\textwidth]{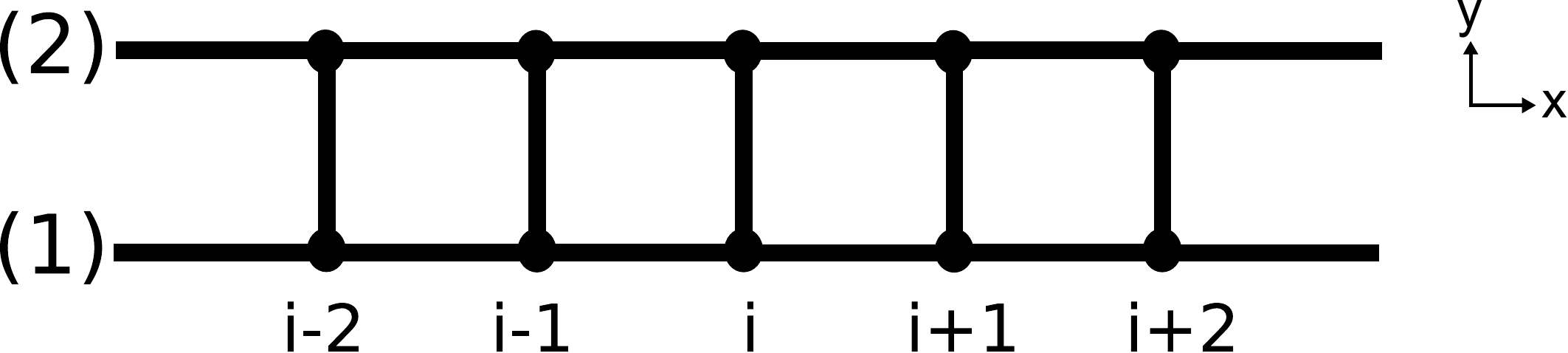}
 \end{center}
 \caption{Two-chain ladder model, where dots denote atom sites and solid lines indicate possible hopping directions.}
 \label{fig:ladder:ladder}
\end{figure}

We consider a two-chain ladder of spinless fermions as depicted in Fig. \ref{fig:ladder:ladder}, with lattice spacing $a_x$ and $a_y$.  The ladder is pierced by a magnetic field in the $z$-direction, and the electrons feel an interaction and a disorder potential.  The total Hamiltonian we will consider is therefore a sum of three parts:
\begin{equation}
 H = H_{\text{kin}} + H_{\text{int},fs} + H_{\text{dis}}.
\end{equation}
We now consider the three terms separately.

\subsection{Kinetic term and magnetic field}

\begin{figure}
\begin{center}
\includegraphics[width=0.35\textwidth]{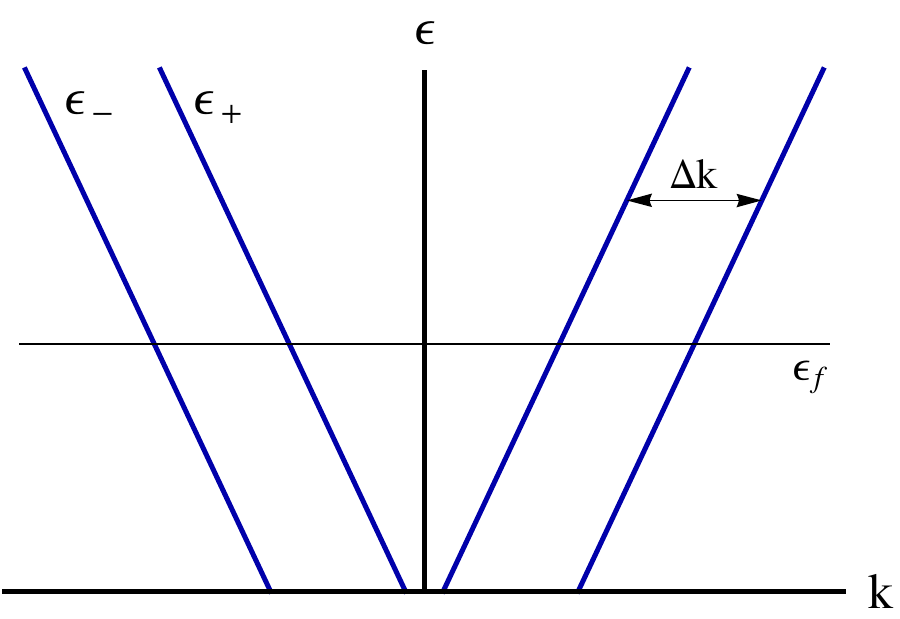}
\end{center}
\caption{Band spectrum of the linearized model. The $+$ and $-$-bands exhibit the same Fermi velocity, but have different Fermi momenta, split by $\Delta k = k_{F,-} - k_{F,+}$.}
\label{fig:ladder:dispersion_dk}
\end{figure}

We take the continuum limit in the $x$ direction, which amounts to linearizing the spectrum; the inter-chain hopping term $t_\perp$ then serves to split the bands (assumed both to have the same Fermi-velocity $v_F$):
\begin{align}\label{eq:ladder:Hkin}
 H_{\text{kin}} = \int dx \biggl[ &\sum_{\mu n} i \mu v_F \Psi^{\dagger}_{\mu n} (x) \partial_x \Psi_{\mu n} (x) \nonumber \\
 &+ t_\perp \sum_{\mu} \left( \Psi^{\dagger}_{\mu 1} (x) \Psi_{\mu 2} (x) + h.c. \right) \biggr].
\end{align}
Here, the $\Psi_{\mu n} (x)$ are the electron operators on the two chains, $n=1,2$ denotes the chain index and $\mu=R,L=+,-$ indicates the chirality (right and left movers).  The Hamiltonian is diagonalized by the transformation $ \Psi_{\mu,\sigma} = (\Psi_{\mu,1} + \sigma \Psi_{\mu,2} )/\sqrt{2} $ with $\sigma = \pm$, which yields the dispersion relations

\begin{equation}\label{eq:ladder:dispersion_nomagnetic}
 \epsilon_{\sigma}^{R}(k) = v_F k + \sigma t_\perp, \qquad \epsilon_{\sigma}^{L}(k) = -(v_F k + \sigma t_\perp).
\end{equation}
The band structure is plotted in Fig. \ref{fig:ladder:dispersion_dk}, note that the only quantitiy which discriminates the $\sigma=\pm$-bands is their respective Fermi momentum $k_{F,\pm}$.

The magnetic field is included in Eq. \eqref{eq:ladder:Hkin} via the vector potential by means of the minimal substitution $\partial_x \rightarrow \partial_x + i e A_x$ in the longitudinal direction; and the Peierls substitution $t_\perp \rightarrow t_\perp e^{ -i e \int \textbf{A} d \textbf{l} }$ in the transverse direction, where the integral is performed over a rung. We apply the magnetic field perpendicular to the ladder, $\textbf{B} = B \textbf{e}_z$ and choose the gauge $\textbf{A}_{1,2} = \mp (B a_y /2) \textbf{e}_x$ for the two chains $1$ and $2$. After Fourier transforming to momentum space, the Hamiltonian is given by

\begin{align}\label{eq:ladder:Hkin_magnetic}
H_{\text{kin}} = \int \frac{dk}{2 \pi} \sum_\mu &\Bigl[ \mu v_F \Bigl( (k+ \pi f) \Psi_{\mu,1}^\dagger(k) \Psi_{\mu,1} (k) \nonumber \\
 & + (k- \pi f)  \Psi_{\mu,2}^\dagger(k) \Psi_{\mu,2} (x) \Bigr)  \nonumber \\
 & + t_\perp \left( \Psi_{\mu,1}^\dagger(k) \Psi_{\mu,2} (k) + h.c.  \right)  \Bigl].
\end{align}
We have introduced the magnetic flux density
\begin{equation}
 f = B a_y / 2 \Phi_0,
\end{equation}
with the magnetic flux quantum $\Phi_0 = h/2e$. Note that $f$ has dimension $[ \text{length}^{-1} ]$, whereas flux is conventionally dimensionless. The reason for this unusual dimension is the linearization of the spectrum, which is equivalent to omitting the lattice spacing in the x-direction. Nevertheless, we take $f$ to be a convenient parameterization of the magnetic field and for convenience will call it magnetic flux.

The Hamiltonian \eqref{eq:ladder:Hkin_magnetic} is diagonalized by the transformation

\begin{align}\label{eq:ladder:transformation}
 \Psi_{\mu,1} &= u_\mu \Psi_{\mu,+} + v_\mu \Psi_{\mu,-}, \nonumber \\ 
 \Psi_{\mu,2} &= v_\mu \Psi_{\mu,+} - u_\mu \Psi_{\mu,-}, 
\end{align}
where
\begin{equation}
 u_R^2 = v_L^2 = \frac{1+b}{2}, \qquad v_R^2 = u_L^2 = \frac{1-b}{2},
\end{equation}
and $b\in[0,1]$ is a dimensionless measure of the strength of the external magnetic field
\begin{equation}\label{eq:ladder:alpha}
b = \frac{v_F \pi f}{ \sqrt{ (v_F\pi f)^2 + t_\perp^2 } }.
\end{equation}
This transformation leads to similar dispersion relations as in Eq. \eqref{eq:ladder:dispersion_nomagnetic}:
\begin{align}
\epsilon^R_\pm (k) &= v_F k \pm \sqrt{ (v_F \pi f)^2 + t_\perp^2 } , \nonumber \\
\epsilon^L_\pm (k) &= -v_F k \mp \sqrt{ (v_F \pi f)^2 + t_\perp^2 }.
\end{align}

The magnetic field affects only the Fermi momenta but leaves the Fermi velocity unchanged. The Fermi momenta splitting changes as 
\begin{equation}\label{eq:ladder:Dk}
k_{F,-} - k_{F,+} = \Delta k = \frac{2}{v_F} \sqrt{ (v_F \pi f)^2 + t_\perp^2  }. 
\end{equation}
Using Eq. \eqref{eq:ladder:Hkin_magnetic}, we also find a mapping between the Green functions in the chain basis and those in the band basis:
\begin{align}\label{eq:ladder:chain-band-mapping}
 G_{\mu}^{(1,1)} &= \frac{1}{2} \left[ \left(1 - \mu b \right) G_{\mu,+} + \left(1 + \mu b \right) G_{\mu,-}  \right], \nonumber \\
 G_{\mu}^{(2,2)} &= \frac{1}{2} \left[ \left(1 + \mu b \right) G_{\mu,+} + \left(1 - \mu b \right) G_{\mu,-}  \right], \nonumber \\
 G_{\mu}^{(1,2)} &= G_{\mu}^{(2,1)} = \frac{1}{2} \sqrt{1 - b^2} \left[ G_{\mu,+} - G_{\mu,-} \right].
\end{align}
Here,
\begin{equation}
G_{\mu}^{(i,j)}(tt';xx') = -i\left< T_t \Psi_{\mu,i}(t,x) \Psi^\dagger_{\mu,j}(t',x') \right>
\end{equation}
are the Green functions for propagation from chain $i$ to chain $j$ whereas $G_{\mu,\pm}$ are the propagators for the single bands. Note that these relations hold only for the linearized model and therefore are valid only for small magnetic fields. Large magnetic fields of order $1/2$ flux quantum per plaquette in the original lattice model affect the curvature of the dispersion relation and can eventually lead to a gap.\cite{Carr-Narozhny-Nersesyan-2006}   Such a field would be unrealistically large for a condensed matter realization of such a model; while this limit may be possible in certain superstructures or optical lattices, this case will not be discussed here.

\subsection{Interactions}
\label{sec:interactions}

We introduce a screened, short-ranged electron-electron interaction via two terms, one describing forward scattering

\begin{equation}
 H_{\text{int},fs} = \frac{1}{2} \sum_{\mu \sigma \sigma'} \int dx \left( n_{\mu \sigma} g_4^{\sigma\sigma'} n_{\mu \sigma'} + n_{\mu \sigma} g_2^{\sigma\sigma'} n_{-\mu \sigma'} \right)
\end{equation}
with the electron density $ n_{\mu \sigma} (x) = \Psi_{\mu \sigma}^{\dagger} (x) \Psi_{\mu \sigma} (x) $, and one term describing backward scattering

\begin{gather}
 H_{\text{int},bs} = \frac{1}{2} \sum_{\mu \sigma \sigma'} \int dx \Psi_{\mu \sigma}^{\dagger} \Psi_{-\mu \sigma} g_1^{\sigma\sigma'} \Psi_{-\mu \sigma'}^{\dagger} \Psi_{\mu \sigma'}
 \nonumber \\
 + \frac{1}{2} \sum_{\mu \sigma \sigma'} \int dx \Psi_{\mu \sigma}^{\dagger} \Psi_{-\mu \sigma}^\dagger \tilde{g}_1^{\sigma\sigma'} \Psi_{-\mu \sigma'} \Psi_{\mu \sigma'}.
\end{gather}
Note that the interaction as shown here is similar to that of a single chain spinful Luttinger liquid. In consequence, the band index $\sigma$ can also be seen as a pseudospin; the unusual looking $\tilde{g}_1$ term (corresponding to spin-flip scattering in the spinful Luttinger liquid analogy) appears as particle number in individual bands is not conserved by the interaction.  Each of the coupling constants can be further split into components acting within the same band ($\sigma=\sigma'$) and different bands ($\sigma=-\sigma'$).

At low temperatures, the interplay between the different interaction terms lead to a rich phase diagram of strong coupling phases,\cite{Nersesyan-Luther-Kusmartsev-1993,Ledermann-LeHur-2000,Carr-Narozhny-Nersesyan-2011} in which the pseudospin excitations acquire a gap.  However as temperature is raised above this gap the system simply behaves as a Luttinger liquid, exhibiting pseduospin-charge separation and power-law correlations.  In this regime, the only role of the back-scattering terms $g_1$ and $\tilde{g}_1$ is to renormalize the forward scattering terms, and so they may be neglected.  Furthermore, in this regime we will neglect the band dependence on the interaction terms and set
\begin{equation}\label{eq:justg}
g_2^{\sigma\sigma'}  = g_4^{\sigma\sigma'}  = g.
\end{equation}
The reason for this is twofold - firstly, these interaction terms correspond to scattering events with low momentum transfer so in any realistic situation they are likely to be similar anyway; secondly and more importantly, the most crucial role of the interaction terms and the consequent Luttinger liquid state for the current work will turn out to be the spin-charge separation, which already occurs in the single parameter model so extra complexity is not needed.

It will be convenient to introduce the dimensionless interaction strength
\begin{equation}\label{eq:dimint}
\alpha=g/\pi v_F.
\end{equation}
As we will show, this one parameter is responsible for both the exponent of renormalization of disorder and dephasing time,\cite{Yashenkin-Gornyi-Mirlin-Polyakov-2008} and will be assumed to be small $\alpha \ll 1$.

\subsection{Disorder}
\label{sec:disorder}

The final part of the Hamiltonian is disorder scattering $ H_{\text{dis}} = \sum_{a=1,2} \int dx U_a(x)n_a(x)$ where the white-noise disorder potential is taken to be uncorrelated between the two chains and chosen from the usual Gaussian distribution
\begin{equation}\label{eq:ladder:impuritycorrelator}
 \left< U_n(x) U_{n'}(x') \right> = \frac{\delta (x-x')\delta_{nn'}}{2 \pi \nu \tau_\mathrm{tr}^{(0)}}.
\end{equation}
Here, $ \nu = (\pi v_F)^{-1} $ is the density of states per chain, and $\tau_\mathrm{tr}^{(0)}$ is the (bare) transport scattering time. The disorder is considered to be weak, $\epsilon_f \tau_\mathrm{tr}^{(0)} \gg 1$. Since forward scattering does not affect transport\cite{Abrikosov-Ryzhkin-1978}, we only consider backward scattering off impurities, so $U(k\sim0) = U_f = 0$ and $U(k \sim 2k_F) = U_b$. Thus we can write the disorder-induced term in the Hamiltonian as
\begin{equation}
 H_{\text{dis}} = \sum_{n=1,2} \int dx \left( U_{b,n}^* \Psi^{\dagger}_{R,n} \Psi_{L,n} + U_{b,n} \Psi^{\dagger}_{L,n} \Psi_{R,n} \right).
\end{equation}
The backscattering amplitudes are correlated as $ \left< U_{b,n}(x) U_{b,n'}(x') \right> = 0 $ and $ \left< U_{b,n}(x) U_{b,n'}^*(x') \right> = \left< U_n(x) U_{n'}(x') \right> $, furthermore we assume that the impurities scatter only intra-chain, but not inter-chain.

At the most elementary level, the disorder gives rise to diffusive behavior of the system and one observes a Drude conductivity
\begin{equation}\label{eq:Drude}
\sigma_D = 2\frac{e^2 v_F}{\pi} \tau_\mathrm{tr},
\end{equation}
however in the interacting system the scattering time is renormalized by the interaction\cite{Giamarchi-Schulz-1988,Gornyi-Mirlin-Polyakov-2007,Yashenkin-Gornyi-Mirlin-Polyakov-2008,Orignac-Giamarchi-1996,Orignac-Giamarchi-1997} and therefore gains a temperature dependence
\begin{equation}\label{eq:taurenorm}
\tau_\mathrm{tr}(T) = \tau_\mathrm{tr}^{(0)} \left( \frac{T}{\Lambda} \right)^\alpha,
\end{equation}
where $\alpha$ is just the dimensionless interaction strength introduced in Eq.~\eqref{eq:dimint}, and $T$ is temperature.

In fact, for the present model, this equation is strictly only valid for the limit $T>v_F \Delta k$ i.e the temperature is greater than the splitting between the bands.  This is because below this temperature, the renormalization of certain inter-band scattering processes will be cut-off by the band splitting and not temperature\cite{Matveev-Yue-Glazman-1993} giving rise to a magnetoresistance effect even at the level of the Drude conductivity.\cite{hurry-up-Andrey}  Furthermore, this band dependence on renormalization will mean that the impurity scattering no longer remains diagonal in the chain basis.  However, for weak interaction this renormalization is weak, and one can ignore this asymmetric renormalization so long as we stay away from the very low temperature $T\ll v_F \Delta k$ limit.  In everything that follows, $\tau_\mathrm{tr}$ is assumed to take its already renormalized, and therefore weakly temperature dependent, value.

Having introduced the Hamiltonian, we now proceed to a calculation of the weak localization correction, which gives the dominant behavior of the magnetoresistance.

\section{Calculation of weak localization correction}
\label{sec:problem}

In the absence of interaction, disorder leads to full localization of all electronic states in one- and two dimensional systems; leading to zero conductivity at any temperature. However, this is no longer the case in the presence of interaction. Inelastic electron-electron scattering leads to a finite dephasing length $l_{\phi}$, which causes a finite weak localization correction to the conductivity, $\Delta \sigma_{WL}$. At high enough temperatures, the correction is small compared to the Drude conductivity $\sigma_D$, but at lower temperatures $ \Delta \sigma_{WL}$ becomes of the order of $\sigma_D$. At this point, the system crosses over to the strong localization regime.

The weak localization correction has previously been calculated for a single-channel spinless Luttinger liquid \cite{Gornyi-Mirlin-Polyakov-2005,Gornyi-Mirlin-Polyakov-2007} and also for the spinful case.\cite{Yashenkin-Gornyi-Mirlin-Polyakov-2008}  We follow a similar approach to calculate the weak localization correction for the spinless disordered two-chain ladder which is layed down in Sec.~\ref{sec:laddermodel}. We restrict the calculation to the limit of strong dephasing, $\tau_\phi/\tau \ll 1$, which is always the case for either sufficiently high temperatures or for sufficiently low disorder.  In this limit, the disorder may be treated perturbatively, and we only have to retain the shortest possible Cooperon, that is, the one with 3 impurity scattering lines.\cite{Gornyi-Mirlin-Polyakov-2007}

\begin{figure}
\begin{center}
\includegraphics[width=0.4\textwidth]{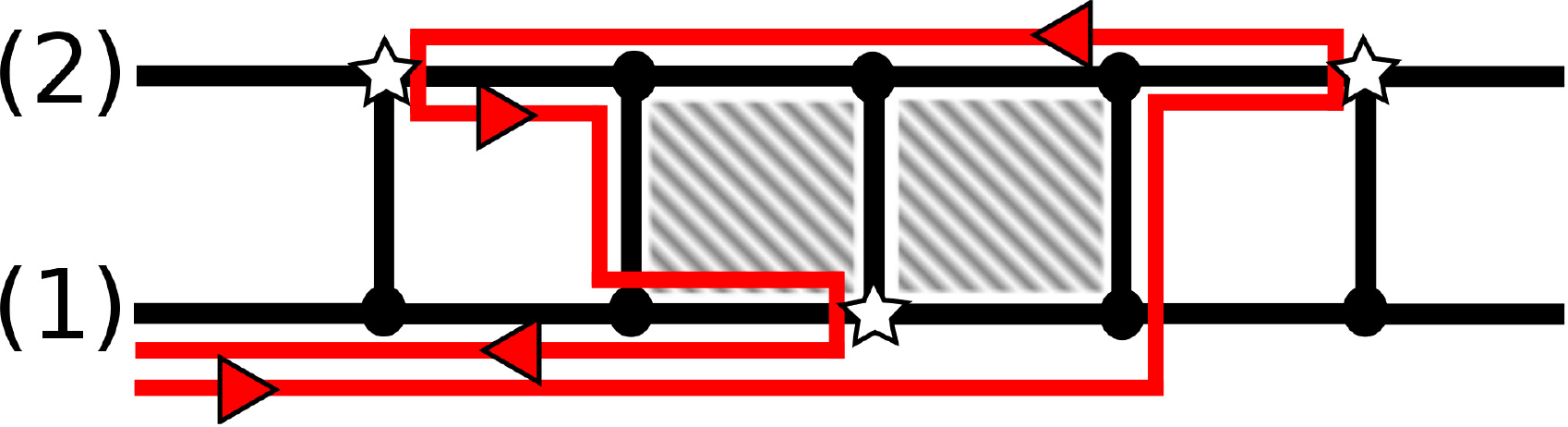}
\end{center}
\caption{Example process for scattering on 3 impurities where a finite area is enclosed. Stars indicate impurities, the red line shows a possible way an electron can take in order to scatter on all three impurities and the shaded region denotes the area which is enclosed by the loop.}
\label{fig:problem:ladder_WL}
\end{figure}

The difference between the pure one-dimensional system and the two-chain ladder is that the electrons can perform loops which enclose a finite area, thus, weak localization is sensitive to a magnetic field, which leads to magnetoresistance. An example of such a process is shown in Fig.~\ref{fig:problem:ladder_WL} for a specific setup of impurity positions. Of course, at the end one should average over all possible impurity configurations.

\begin{figure}
\begin{center}
\includegraphics[width=0.45\textwidth]{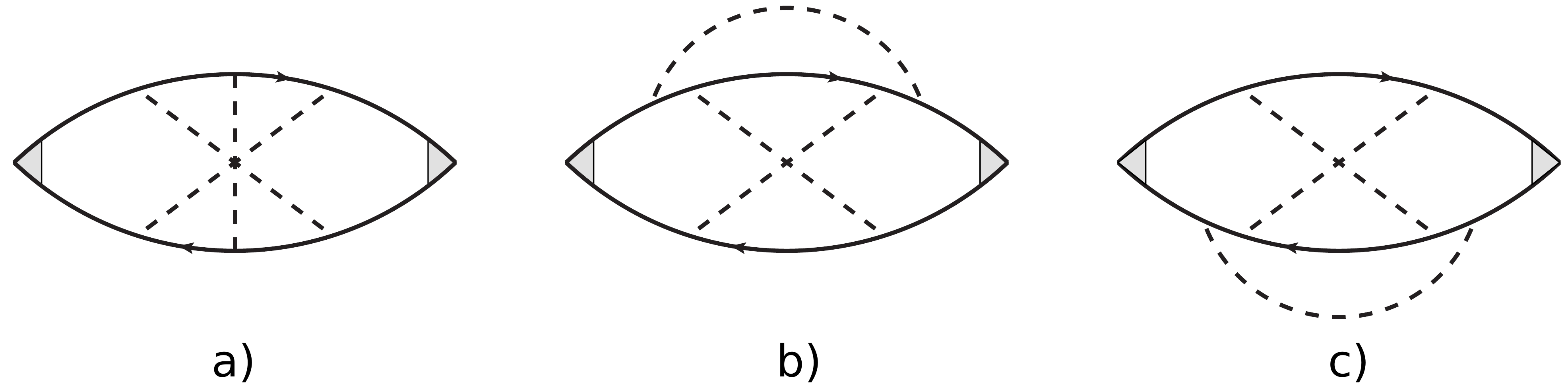}
\end{center}
\caption{Leading order diagrams to the weak localization correction in the limit $\tau_\phi/\tau \ll 1$. Dashed lines describe backscattering off impurities, the current vertices are dressed with diffusons and the solid lines are Green functions with disorder-induced self-energies. The diagrams are understood as fully dressed by electron-electron interactions.}
\label{fig:problem:phen:diagrams}
\end{figure}

The weak localization correction depicted schematically in Fig.~\ref{fig:problem:ladder_WL} corresponds formally to calculating the diagrams in Fig.~\ref{fig:problem:phen:diagrams}, the essential point being that the diagrams should be fully dressed by interactions.  We now calculate this in two different ways -- firstly in Sec.~\ref{sec:problem:phen} we treat the interactions phenomenologically assuming that their only role is to introduce a dephasing time $\tau_\phi$ into the problem which is introduced by hand.  The advantage of this approach is transparency -- one ends up with an analytic answer and can understand the different geometrical processes that give rise to different terms in this answer.  Secondly in Sec.~\ref{sec:problem:micro}, we calculate the diagram within a full microscopic approach by employing the technique of functional bosonization.\cite{Yurkevich-2001}  By matching the phenomenological approach with the microscopic one in Sec.~\ref{sec:problem:comparison} we show that the former calculation is for the most part a very good approximation, and we find an explicit expression, Eq.~\eqref{tauphiint}, for the relation between the dephasing time $\tau_\phi$ and the interaction strength $\alpha$.  We also discuss an effect that goes beyond the phenomenological picture and depends in a more essential way on the nature of the correlations in the Luttinger liquid  -- namely magnetoresistance oscillations.

\subsection{Phenomenological approach}
\label{sec:problem:phen}

We start by calculating the weak localization correction in a phenomenological way, closely following the method presented in Appendix B of Ref.~\onlinecite{Gornyi-Mirlin-Polyakov-2007}.  The basic idea is that one considers the retarded Green function $G_{R/L,\sigma}(\omega,k)$ to be of the non-interacting form (but with a renormalized scattering time, as discussed in Sec.~\ref{sec:disorder}):
\begin{equation}\label{eq:problem:phen:GF}
 G_{R/L,\sigma}(\omega,k) =   \left[ \omega - v_F (\pm k - k_{F,\sigma} ) + i/4\tau_\mathrm{tr}  \right]^{-1};
 \end{equation}
 the numerical factor in $1/4\tau_\mathrm{tr}$ comes from the specifics of the model with only backscattering, and expresses the total scattering rate in terms of the transport scattering rate.
One then adds a phenomenological dephasing time $\tau_\phi$ introduced as an additional source of decay via the substitution
\begin{equation}\label{eq:problem:phen:GF-dephasing}
 G_{R/L,\sigma}(\omega,k)  \rightarrow   \left[ \omega - v_F (\pm k - k_{F,\sigma} ) + i/4\tau_\mathrm{tr} + i/2\tau_\phi \right]^{-1}.
\end{equation}
The crucial difference between the transport lifetime $\tau_\textrm{tr}$ and the dephasing time $\tau_\phi$ is that the latter should be included only for Green functions running \textit{within the weak localization loop}.\cite{Gornyi-Mirlin-Polyakov-2007}  In this sense, the inverse dephasing time which accounts for all inelastic processes acts as an effective lower cutoff on allowed momenta in the weak localization correction, as it should.\cite{Gorkov-Larkin-Khmelnitskii-1979}

The leading order diagrams for the weak localization correction in the limit $\tau_\phi / \tau \ll 1$ are shown in Fig. \ref{fig:problem:phen:diagrams}. 
The two diagrams \ref{fig:problem:phen:diagrams}b and \ref{fig:problem:phen:diagrams}c sum up to the value of diagram \ref{fig:problem:phen:diagrams}a, so the total correction is given by

\begin{equation}
 \Delta \sigma_{WL}^{\text{phen}} = 2 \sigma_{C3}.
\end{equation}
$\sigma_{C3}$ is the fully crossed diagram \ref{fig:problem:phen:diagrams}a and is given by
\begin{equation}\label{eq:problem:phen:sigmaC3}
 \sigma_{C3} = -2 \left( e v_F \right)^2 \int \frac{d \epsilon}{2 \pi} \left( - \frac{\partial f_{\epsilon}}{\partial \epsilon} \right) \int \frac{dQ}{2\pi} J(Q),
\end{equation}
where the $2$ comes from the two possibilities to choose the chiralities at the current vertices, the minus sign reflects the different signs of the two current vertices and $f_{\epsilon}$ is the Fermi distribution. The function $J(Q)$ is given by
\begin{align}
 J(Q) &= \frac{1}{(2 \pi \nu \tau_\mathrm{tr})^3} \sum_{ijklm} \int \frac{dp}{2\pi} \int \frac{dp_1}{2\pi} \int \frac{dp_2}{2\pi} \nonumber \\
      &\times G^{(i,j)}_{R}(p) G^{(j,k)}_{L}(p_1) G^{(k,l)}_{R}(p_2) G^{(l,m)}_{L}(-p+Q) \nonumber \\
      &\times \Bigl[ G^{(i,l)}_{R}(p) G^{(l,k)}_{L}(-p_2+Q) \nonumber \\
      &\times  G^{(k,j)}_{R}(-p_1+Q) G^{(j,m)}_{L}(-p+Q) \Bigr]^*,
\end{align}
where we sum over all possible chain setups of the impurities. From Eq. \eqref{eq:ladder:chain-band-mapping} we can transform the Green functions into the band representations and find
\begin{align}\label{eq:problem:phen:Jband}
 J(Q) &= \frac{1}{(2 \pi \nu \tau_\mathrm{tr})^3} \sum_{ijkli'j'k'l'} P^{ijkl}_{i'j'k'l'} \int \frac{dp}{2\pi} \int \frac{dp_1}{2\pi} \int \frac{dp_2}{2\pi} \nonumber \\
      &\times G_{R,i}(p) G_{L,j}(p_1) G_{R,k}(p_2) G_{L,l}(-p+Q) \nonumber \\
      &\times \Bigl[ G_{R,i'}(p) G_{L,j'}(-p_2+Q) \nonumber \\
      &\times  G_{R,k'}(-p_1+Q) G_{L,l'}(-p+Q) \Bigr]^*.
\end{align}
The sum over the different chains as well as the prefactors in Eq.~\eqref{eq:ladder:chain-band-mapping} have been absorbed in $P^{ijkl}_{i'j'k'l'}$, the sum in Eq.~\eqref{eq:problem:phen:Jband} is over the different band setups. The exact definition of $P^{ijkl}_{i'j'k'l'}$ is given in Table \ref{tab:problem:phen:P}.

\begin{table}[t]
\begin{center}
 \begin{tabular}{l|l|l}
$P^{ \{ \gamma \} }_{ \{ \delta \} }$ & $ \{ \gamma \} $ & $ \{ \delta \} $\\ \hline \hline
$ (1 - b^2)^3 /8 $ & $ \{ i,j,k,i \} $ & $ \{ i,k,j,i \} $\\ \hline
$ (1 + b^2)^3 /8 $ & $ \{ i,-i,i,-i \} $ & $ \{ i,-i,i,-i \} $\\ \hline
$ (1 + b^2) (1 - b^2)^2 /8 $ & $ \{ i,j,j,-i \} $ & $ \{ i,-j,-j,-i \} $\\
  & $ \{ i,i,-i,-i \} $ & $ \{ i,i,-i,-i \} $\\ \hline
$ (b^2 - b^6) /8 $ & $ \{ i,j,-j,i \} $ & $ \{ i,j,-j,i \} $\\
  & $ \{ i,j,i,-i \} $ & $ \{ i,-i,j,-i \} $\\ \hline
$ -b^2 (1 - b^2)^2 /8 $ & $ \{ i,j,j,i \} $ & $ \{ i,-j,-j,i \} $\\
  & $ \{ i,i,-i,-i \} $ & $ \{ i,j,j,-i \} $\\\hline
$ b^2 (1 - b^2)^2 /8 $ & $ \{ i,j,k,-i \} $ & $ \{ i,k,j,-i \} $\\
  & $ \{ i,j,j,i \} $ & $ \{ i,j,-j,i \} $\\
  & $ \{ i,j,j,i \} $ & $ \{ i,-j,j,i \} $
  \end{tabular}
\end{center}
\caption{Values of the factors $P^{ \{ \gamma \} }_{ \{ \delta \} }$ for a given set of indices $ \{ \gamma \} $ and $ \{ \delta \} $. All latin indices can exhibit the values $+$ and $-$. Combinations not listed here result in $P^{ \{ \gamma \} }_{ \{ \delta \} }=0$. The two sets $ \{ \gamma \} $ and $ \{ \delta \} $ can be switched without changing the value of $P^{ \{ \gamma \} }_{ \{ \delta \} }$.}
\label{tab:problem:phen:P}
\end{table}

The current vertices should be dressed by the insertion of diffusons; for the backscattering only model this amounts simply to the replacement $e v_F \rightarrow e v_F /2$.  Furthermore, the Green functions connected to the current vertices can be decomposed, 
\begin{align}\label{eq:problem:phen:GF-Decomposition}
 &G_{\mu,\sigma} (p) \left( G_{\mu,\sigma'} (p) \right)^*  \nonumber \\
 &= \frac{2 i \tau_\textrm{tr}}{1 - 2 i \tau_\textrm{tr} v_F [ k_{F,\sigma} - k_{F,\sigma'} ]} \left[ G_{\mu,\sigma} (p) -  \left( G_{\mu,\sigma'} (p) \right)^* \right],
\end{align}
which leads to
\begin{align}\label{eq:problem:phen:JbandF}
 J(Q) = \sum_{ijkli'j'k'l'} &\tilde{P}^{ijkl}_{i'j'k'l'} F^{LR}_{jk'} (Q) \left( F^{LR}_{j'k}(Q) \right)^* \nonumber \\
 &\times \left[ F^{LR}_{i'l} (Q) + \left( F^{LR}_{il'}(Q) \right)^*  \right].
\end{align}
The prefactors of Eq.~\eqref{eq:problem:phen:GF-Decomposition} have been absorbed in the definition of $\tilde{P}^{ijkl}_{i'j'k'l'}$ and the functions $F(Q)$ are given by
\begin{equation}
 F^{pq}_{ij} (Q) = \frac{1}{2 \pi \nu \tau_\mathrm{tr}} \int \frac{dp}{2\pi} G_{p,i}(p) \left( G_{q,j} (Q-p) \right)^*
\end{equation}
and $ F^{pq}_{ij} = \left( F^{qp}_{ji} \right)^* $. 

Note that Eq.~\eqref{eq:problem:phen:JbandF} describes only the loop between the impurities, the diffusive motion has already been taken care of by Eq. \eqref{eq:problem:phen:GF-Decomposition}. Consequently, it is at this stage we make the substitution \eqref{eq:problem:phen:GF-dephasing} and include the phenomonological dephasing time in the Green functions.  Hence the $F(Q)$ functions may be evaluated
\begin{align}\label{eq:problem:phen:Ffunctions}
F_{++}^{LR} (Q) = F_{--}^{LR} (Q) &= \frac{1}{2\tau_\mathrm{tr}} \frac{1}{ 1/2\tau_\mathrm{tr} + 1/\tau_\phi + i v_F Q },  \\
F_{+-}^{LR} (Q) &= \frac{1}{2 \tau_\mathrm{tr}} \frac{1}{1/2\tau_\mathrm{tr} + 1/\tau_\phi + i v_F ( Q -\Delta k) }, \nonumber \\
F_{-+}^{LR} (Q) &= \frac{1}{2 \tau_\mathrm{tr}} \frac{1}{1/2\tau_\mathrm{tr} + 1/\tau_\phi + i v_F ( Q + \Delta k)}, \nonumber
\end{align}
where $\Delta k = k_{F,-} - k_{F,+}$. 

Combining \eqref{eq:problem:phen:Ffunctions}, \eqref{eq:problem:phen:JbandF} and \eqref{eq:problem:phen:sigmaC3} we therefore find for the weak localization correction to the conductivity in the limit $\tau_\phi \ll \tau_\mathrm{tr}$ to be
\begin{equation}\label{eq:problem:phen:WLcorrection}
 \Delta \sigma_{WL}^{\text{phen}} = -\frac{1}{8} \sigma^{(1)}_D \left( \frac{\tau_\phi}{\tau_\mathrm{tr}} \right)^2 K_{\text{phen}} (b, v_F \tau_\phi \Delta k).
\end{equation}
Here, $\sigma^{(1)}_D = e^2 v_F \tau_\mathrm{tr} / \pi$ is Drude conductivity of a single chain -- this is introduced for convenience, the actual Drude conductivity of the two-leg ladder \eqref{eq:Drude} is simply given by $\sigma_D=2  \sigma^{(1)}_D$.
The dimensionless function $K_{\text{phen}}$ contains the interesting functional dependence of the weak localization correction and is given by
 \begin{align}\label{eq:problem:phen:Kphen}
K_{\text{phen}}(b, v_F &\tau_\phi \Delta k) = \nonumber \\
= \frac{1}{4} \Biggl\{ b^0 \Biggl[ 5 &+ \frac{1}{1+(v_F \tau_\phi \Delta k)^2}+\frac{2}{\left(1+(v_F \tau_\phi \Delta k)^2\right)^2} \Biggr] \nonumber \\
 + b ^2 \Biggl[-9 &+ \frac{-9}{1+(v_F \tau_\phi \Delta k)^2}+\frac{-2}{\left(1+(v_F \tau_\phi \Delta k)^2\right)^2} \nonumber \\
&+\frac{32}{4+(v_F \tau_\phi \Delta k)^2}+\frac{192}{\left(4+(v_F \tau_\phi \Delta k)^2\right)^2}\Biggr] \nonumber\\
+b^4 \Biggl[15 &+ \frac{15}{1+(v_F \tau_\phi \Delta k)^2}+\frac{-2}{\left(1+(v_F \tau_\phi \Delta k)^2\right)^2} \nonumber \\
&+\frac{-48}{4+(v_F \tau_\phi \Delta k)^2}+\frac{-256}{\left(4+(v_F \tau_\phi \Delta k)^2\right)^2}\Biggr] \nonumber\\
+b^6 \Biggl[-3 &+ \frac{-7}{1+(v_F \tau_\phi \Delta k)^2}+\frac{2}{\left(1+(v_F \tau_\phi \Delta k)^2\right)^2} \nonumber \\
&+\frac{16}{4+(v_F \tau_\phi \Delta k)^2}+\frac{64}{\left(4+(v_F \tau_\phi \Delta k)^2\right)^2}\Biggr] \Biggr\},
\end{align}
with $b$ defined in Eq.~\eqref{eq:ladder:alpha}. The properties of this result will be examined in detail in Sec.~\ref{sec:discussion}, but first we will compare this with a direct microscopic calculation.  This will allow us both to relate the dephasing time introduced here to the microscopic parameters of the model as well as to establish limits on the validity of the above formula.


\subsection{Microscopic approach}
\label{sec:problem:micro}

Having treated the electron-electron interaction qualitatively in the last section via a phenomenological dephasing time, we now turn to a true microscopic calculation using the method of functional bosonization. This method was introduced in Refs. \onlinecite{Fogedby-1976} and \onlinecite{Lee-Chen-1988} for the case of a clean Luttinger liquid and further developed in Refs.~\onlinecite{Yurkevich-2001,Grishin-Yurkevich-Lerner-2004,Kopietz-1997,Naon}. The framework has recently also been extended to the disordered case, where the transport properties of a spinless \cite{Gornyi-Mirlin-Polyakov-2005,Gornyi-Mirlin-Polyakov-2007} and spinful \cite{Yashenkin-Gornyi-Mirlin-Polyakov-2008} disordered Luttinger liquid have been studied. Functional bosonization has the advantage that it preserves both fermionic and bosonic degrees of freedom. This property is extremely helpful in the case of disordered Luttinger liquids, since interaction and disorder can be treated on an equal footing.  Here, we extend the previous work on single (but possibly spinful) channel Luttinger liquids to the case of a spinless two-leg ladder.  This extension is simplified by remembering that the two chains may be thought of as pseudo-spins, which allows us closely to follow the formalism developed in Ref.~\onlinecite{Yashenkin-Gornyi-Mirlin-Polyakov-2008}.

The basic principles of functional bosonization as applied to disorder diagrams are summarized in Appendix \ref{sec:appendix:funcbos}.  In essence, one factorizes the full Green function $G_{\mu}(x,\tau)$ in real space into a non-interacting part $g_\mu^0(x,\tau)$ and an exponent of an interaction (bosonic) correlator, $B_{\mu \mu}(x,\tau)$:
\begin{equation}
 G_{\mu} (x,\tau) = g^{0}_{\mu} (x,\tau) \exp \left[ -B_{\mu \mu}(x,\tau) \right].
\end{equation}
Here, $\mu$ is the chirality, and the interaction correlator $B_{\mu \mu}(x,\tau)$ which is given by Eq.~\eqref{eq:appendix:funcbos:B_generic_realspace}.  The crucial point now is that in our interaction model Eq.~\eqref{eq:justg}, only forward scattering terms with zero-momentum transfer are retained.  Hence all information about the original band structure of the ladder model -- that is, the splitting of the bands by the transverse hopping and the external magnetic field -- is encoded in the free Green function $g_\mu^0(x,\tau)$, leaving the interaction propagators unaffected by the geometric details.   The splitting of the two bands in the two-leg ladder (giving each band its own Fermi momentum $k_{F,\sigma}$) is accounted for by an additional phase factor in the free Green function:
\begin{equation}\label{eq:problem:micro:freeGFband}
 g^{0}_{\mu,\sigma} (x,\tau) \rightarrow e^{i \mu k_{F,\sigma} x} g^{0}_{\mu} (x,\tau),
\end{equation}
where $g^{0}_{\mu}(x,\tau)$ is the free Green function of the single channel Luttinger liquid, given by Eq.~\eqref{eq:freegee} in Appendix \ref{sec:appendix:funcbos}.  The propagators in the chain basis are then given by the usual linear combinations, Eq.~\eqref{eq:ladder:chain-band-mapping}.  Hence the only difference between the present situation of a two-leg ladder (in a magnetic flux) and the single chain with spin is the addition of linear combinations of these phase factors.  This property allows for an easy adaption of the methods in Ref.~\onlinecite{Yashenkin-Gornyi-Mirlin-Polyakov-2008}; we therefore outline the calculation referring to the original paper for full details.

When calculating diagrams within the functional bosonization approach, the relevant interaction propagators must be added between every pair of vertices (see Appendix \ref{sec:diagrams}).  For observable quantities (which are given by closed fermionic loops), this general principle amounts to retaining factors of
\begin{equation}\label{eq:problem:micro:Q}
 Q (x,\tau) = \exp \left[ B_{RR}(x,\tau) + B_{LL}(x,\tau) - 2 B_{RL}(x,\tau) \right].
\end{equation}
 The $Q(x,\tau)$ functions stem from the disorder, every pair of backscattering vertices at points $(x_N,\tau_N)$ and $(x_{N'},\tau_{N'})$ is accounted by a factor $Q(x,\tau)$ (with $x=x_{N} - x_{N'}$, $\tau = \tau_{N} - \tau_{N'}$) if the chiralities of the incident electrons are the same, and by a factor of $Q^{-1}(x,\tau)$ if they are different.

\begin{figure}
\begin{center}
\includegraphics[width=0.35\textwidth]{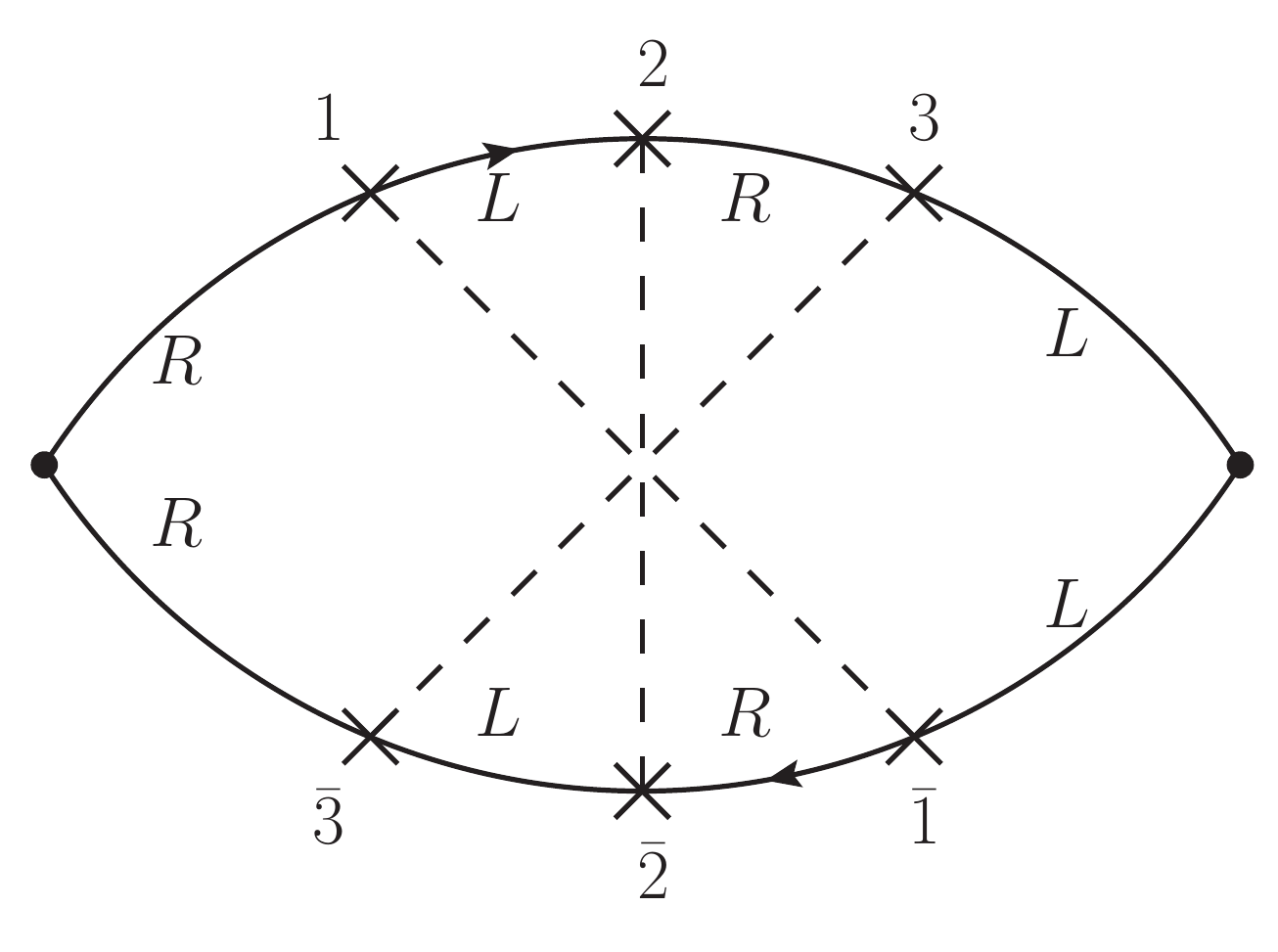}
\end{center}
\caption{Three-impurity Cooperon, crosses denote impurities, solid lines represent free Green functions. The space-time points of the impurities are denoted by $N=(x_N,\tau_N)$ and $\bar{N} = (x_{N} ,\tau_{\bar{N}} )$. Each pair of backscattering vertices has to be accounted by a factor of $Q$ or $Q^{-1}$ (see text).}
\label{fig:problem:micro:cooperon}
\end{figure}

Using these relations, the leading order weak localization correction in the functional bosonization scheme is given by
\begin{widetext}
 \begin{align}\label{eq:problem:micro:WLcorrection_start}
   \Delta\sigma_{WL}^{\text{micro}} &= \lim_{\Omega \rightarrow 0} \Biggl \{ 4 (ev_F)^2 \left( \frac{v_F^2}{2 l} \right)^3 \frac{1}{\Omega_m} \frac{T}{L}  \sum_{ijkli'j'k'l'} P^{ijkl}_{i'j'k'l'} \int_0^{1/T} d \tau_1 d \bar{\tau}_1 d \tau_2  d \bar{\tau}_2  d \tau_3  d \bar{\tau}_3 \int dx_1 dx_2 dx_3 \nonumber \\
 & \times [ g^0_{R} (x_1 - x_3, \tau_1 - \bar{\tau}_3) Q^{-1}(x_1 - x_3, \tau_1 - \bar{\tau}_3) ] [ g^0_{L,j} (x_2 - x_1, \tau_2 - \tau_1) Q^{-1}(x_2 - x_1, \tau_2 - \tau_1) ] \nonumber \\
 & \times [ g^0_{R,k} (x_3 - x_2, \tau_3 - \tau_2) Q^{-1}(x_3 - x_2, \tau_3 - \tau_2) ] [ g^0_{L} (x_1 - x_3, \bar{\tau}_1 - \tau_3) Q^{-1}(x_1 - x_3, \bar{\tau}_1 - \tau_3) ] \nonumber \\
 & \times [ g^0_{R,k'} (x_2 - x_1, \bar{\tau}_2 - \bar{\tau}_1) Q^{-1}(x_2 - x_1, \bar{\tau}_2 - \bar{\tau}_1) ] [ g^0_{L,j'} (x_3 - x_2, \bar{\tau}_3 - \bar{\tau}_2) Q^{-1}(x_3 - x_2, \bar{\tau}_3 - \bar{\tau}_2) ] \nonumber \\
 & \times Q (x_1 - x_3, \tau_1 - \tau_3) Q (x_1 - x_3, \bar{\tau}_1 - \bar{\tau}_3) Q (x_1 - x_2, \tau_1 - \bar{\tau}_2) Q (x_2 - x_1, \tau_2 - \bar{\tau}_1)\nonumber \\
 & \times Q (x_3 - x_2, \tau_3 - \bar{\tau}_2) Q (x_2 - x_3, \tau_2 - \bar{\tau}_3) Q^{-1} (0, \tau_1 - \bar{\tau}_1) Q^{-1} (0, \tau_2 - \bar{\tau}_2)  \nonumber \\
 & \times Q^{-1} (0, \tau_3 - \bar{\tau}_3) \mathcal{W}^{in,R}_{i,i'} ( x_1 - x_3, \tau_1, \bar{\tau}_3, \Omega_m ) \mathcal{W}^{out,L}_{l,l'} ( x_1 - x_3, \bar{\tau}_1, \tau_3, \Omega_m ) \Biggr \}_{i \Omega_m \rightarrow \Omega + i 0}  ,
\end{align}
\end{widetext}
where the space-time points correspond to the ones in Fig.~\ref{fig:problem:micro:cooperon}, the factor $4 = 2 \times 2$ comes from the 2 different choices to choose the chirality at the current vertices and the summation over diagrams (see Fig.~\ref{fig:problem:phen:diagrams}), $L$ is the system size and the $P^{ijkl}_{i'j'k'l'}$ are the same geometric factors as for the phenomenological case, given in Table \ref{tab:problem:phen:P}. The factor $v_F^2/2l$ stems from the impurity correlator defined in eq. \eqref{eq:ladder:impuritycorrelator}, using the (renormalized) mean free path $l=v_F\tau_\mathrm{tr}$.

The factors $\mathcal{W}^{in/out,R/L}_{i,j}(x,\tau,\tau',\Omega_m)$ come from the integration over external coordinates and times of the two Green functions attached to the current vertices [compare to Eq.~\eqref{eq:problem:phen:GF-Decomposition} in the phenomenological picture], and are defined as
\begin{align}
 \int &dx_i d\tau_i e^{-i \Omega_m \tau_i} e^{- \left[ |x_{\alpha} - x_i| + |x_{\beta} - x_i| \right] / l } \times \nonumber \\
 &\times g^{0}_{\mu,i}(x_{\alpha} - x_{i}, \tau_{\alpha} - \tau_{i}) g^{0}_{\mu,j}(x_{i} - x_{\beta}, \tau_{i} - \tau_{\beta}) \nonumber \\
=&g^{0}_{\mu}(\mu(x_{\alpha}-x_{\beta}), \tau_{\alpha}-\tau_{\beta})\mathcal{W}^{in,\mu}_{i,j} (x_\alpha - x_\beta, \tau_\alpha, \tau_\beta, \Omega_m ),
\end{align}
and
\begin{equation}
 \mathcal{W}^{out,\mu}_{i,j}(x,\tau_\alpha,\tau_\beta,\Omega_m) = \mathcal{W}^{in,\mu}_{i,j}(x,\tau_\alpha,\tau_\beta,-\Omega_m).
\end{equation}
While it is not difficult to evaluate the full expression for $\mathcal{W}^{in/out,R/L}_{i,j}(x,\tau,\tau',\Omega_m)$ (see Ref.~\onlinecite{Yashenkin-Gornyi-Mirlin-Polyakov-2008}), it can be immediately simplified by noticing two things.  Firstly, all summands in Eq.~\eqref{eq:problem:micro:WLcorrection_start} where $i \neq i' $ or $ l \neq l' $ vanish (see Table \ref{tab:problem:phen:P}), so only the $\mathcal{W}^{in/out,R/L}_{i,j}(x,\tau,\tau',\Omega_m)$ with the same band indices survive.  Secondly, in the strong dephasing limit which we are considering, only the leading order expansion in $x/l$ is needed; longer loops being exponentially supressed.  Finally, by including the vertex corrections (which amounts to replacement of the total scattering rate $2 v_F / l$ by the transport rate $v_F / l$) we find that\cite{Yashenkin-Gornyi-Mirlin-Polyakov-2008}
\begin{align}\label{eq:problem:micro:Wfunction_sameband}
 \mathcal{W}^{in,\mu}_{i,i}&( x, \tau_\alpha, \tau_\beta, \Omega_m ) = \nonumber \\
 &=-i e^{\mu i k_{F,i} x} \frac{\sgn(\Omega_m)}{|\Omega_m| + v_{F}/l} \left( e^{-i \Omega_m \tau_{\alpha}} - e^{-i \Omega_m \tau_{\beta}} \right).
\end{align}

From Eqs.~\eqref{eq:problem:micro:freeGFband} and \eqref{eq:problem:micro:Wfunction_sameband} we find that the sum in Eq.~ \eqref{eq:problem:micro:WLcorrection_start} simply consists of phase factors $e^{i k_{F,i} x}$ and the factors $P^{ijkl}_{i'j'k'l'}$. Performing the sum results in the function
\begin{widetext}
\begin{align}\label{eq:gammaWL}
 \Gamma_{WL} (b, \Delta k x_a,  \Delta k x_b,  \Delta k x_c)
  =\frac{1}{4} \Bigl\{ b^0 \bigl[ &5 + \cos(2 \Delta k x_a) + \cos(2 \Delta k x_b) + \cos(2 \Delta k x_c) \bigr] \nonumber\\
 +b^2 \bigl[ &-9 + 6\bigl( \cos( \Delta k x_a) + \cos( \Delta k x_b) + \cos( \Delta k x_c) \bigr)\nonumber \\ 
 & -1\bigl( \cos(2 \Delta k x_a) + \cos(2 \Delta k x_b) + \cos(2 \Delta k x_c) \bigr)\nonumber \\
 & -2 \bigl( \cos( \Delta k( x_a-x_b)) + \cos(\Delta k ( x_a+x_c)) + \cos(\Delta k ( x_b+x_c)) \bigr) \bigr]\nonumber \\
 + b^4 \bigl[ &15 - 8\bigl( \cos(\Delta k x_a) + \cos(\Delta k x_b) + \cos(\Delta k x_c) \bigr)\nonumber \\ 
 & -1 \bigl( \cos(2 \Delta k x_a) + \cos(2 \Delta k x_b) + \cos(2 \Delta k x_c) \bigr)\nonumber \\
 & +4 \bigl( \cos(\Delta k( x_a-x_b)) + \cos(\Delta k (x_a+x_c)) + \cos(\Delta k (x_b+x_c)) \bigr) \bigr] \nonumber \\
 + b^6 \bigl[ & -3 + \left(\cos(2 \Delta k x_a) + \cos(2 \Delta k x_b) + \cos(2 \Delta k x_c)\right) \nonumber \\
 & +2 \bigl( \cos(\Delta k( x_c-x_a)) + \cos(\Delta k (x_a+x_b)) + \cos(\Delta k (x_c-x_b)) \bigr) \nonumber \\
 & -2 \bigl( \cos(\Delta k (x_a-x_b)) + \cos(\Delta k (x_a+x_c)) + \cos(\Delta k (x_b+x_c)) \bigr) \bigr] \Bigr\},
\end{align}
\end{widetext}
where $\Delta k = k_{F,-} - k_{F,+}$ is the splitting of the Fermi points, $b$ is given in Eq. \eqref{eq:ladder:alpha} and the new variables
\begin{align}
 x_a &= x_1 - x_3, \nonumber \\
 x_b &= x_3 - x_2, \nonumber \\
 x_c &= x_1 - x_2,
\end{align}
which clearly satisfy $x_a+x_b=x_c$.

The remainder of Eq.~\eqref{eq:problem:micro:WLcorrection_start} (i.e. everything which carries no band index) is identical to the expression for the weak localization correction of the spinful Luttinger liquid (apart from a factor of 2 from spin degeneracy which now appears in the geometrical factor $\Gamma_{WL}$), which was solved in Ref. \onlinecite{Yashenkin-Gornyi-Mirlin-Polyakov-2008}.  The essence of the calculation (see Ref.~\onlinecite{Yashenkin-Gornyi-Mirlin-Polyakov-2008} for details) relies on noticing that in the weakly interacting limit $\alpha\ll 1$, most of the $Q$ factors cancel, the only interaction correlators remaining being those corresponding to the Green functions themselves.  Putting all of these factors together, we find that the weak localization correction is given by
\begin{gather}
\frac{\Delta\sigma_{WL}}{\sigma^{(1)}_{D}} =  \lim_{\Omega \rightarrow 0} \biggl\{ - \frac{2 \pi T}{\Omega_m} \frac{v^4}{32 l^4} \frac{v^4}{(|\Omega_m|+v/l)^2}  \sum_n \nonumber \\
\times  \int_0^{1/T} d\tau_a\bar{\tau}_a d\tau_b\bar{\tau}_b d\tau_c\bar{\tau}_c   \int_0^\infty dx_a dx_b dx_c \delta (x_a + x_b - x_c) \nonumber \\
\times \exp [ i(\Omega_n+\epsilon_n)(\tau_a+\tau_b+\tau_c) + i\epsilon_n (\bar{\tau}_a+\bar{\tau}_b+\bar{\tau}_c)] \nonumber \\
\times \Gamma_{WL}(b, \Delta k x_a,\Delta k  x_b,\Delta k  x_c) G_R(x_a,\tau_a) G_L(x_a,\bar{\tau}_a)   \nonumber \\
\times G_R(x_b,\tau_b) G_L(x_b,\bar{\tau}_b) G_R(x_c,\tau_c) G_L(x_c,\bar{\tau}_c) \biggr\}_{i \Omega_m \rightarrow \Omega + i 0},
\end{gather}
where $G_\mu$ without band indices refer to the full interacting Green functions but without the fast oscillating $e^{ik_F x}$ factors; and are given in Eq.~\eqref{eq:GFs1} in Appendix \ref{sec:GFs}.  This is conveniently written in terms of the Green functions in the mixed $(x,\epsilon)$ representation (see Appendix \ref{sec:GFs}), which gives:
\begin{align}
 \frac{\Delta\sigma_{WL}}{\sigma^{(1)}_{D}} = & \lim_{\Omega \rightarrow 0} \biggl\{ - \frac{2 \pi T}{\Omega_m} \frac{v^4}{32 l^4} \frac{v^4}{(|\Omega_m|+v/l)^2} \nonumber \\
& \times \sum_n \int_0^\infty dx_a dx_b dx_c \delta (x_a + x_b - x_c) \nonumber \\
& \times G^r_+ (x_a,i\epsilon_n + i \Omega_m) G^r_+ (x_b,i\epsilon_n + i \Omega_m) \nonumber \\ 
& \times G^r_+ (x_c,i\epsilon_n + i \Omega_m) G^a_- (x_a,i\epsilon_n) \nonumber \\
& \times G^a_- (x_b,i\epsilon_n) G^a_- (x_c,i\epsilon_n) \nonumber \\
& \times \Gamma_{WL}(b, \Delta k x_a,\Delta k  x_b,\Delta k  x_c) \biggr\}_{i \Omega_m \rightarrow \Omega + i 0},
\end{align}
where $G^r_+$ and $G^a_-$ are given in Eq.~\eqref{eq:defGar}.

Finally, performing the analytical continuation $i \Omega_m \rightarrow \Omega + i 0$, we find in the DC limit $\Omega \rightarrow 0$
\begin{equation}\label{eq:problem:micro:WLcorrection_final}
 \Delta \sigma_{WL}^{\text{micro}} = - \frac{1}{8} \sigma^{(1)}_{D} \left( \frac{l_{ee}}{l} \right)^2 K_{\text{micro}}\left( b, \frac{\Delta k l_{ee}}{2} \right),
\end{equation}
where $l_{ee} \simeq v_{F} / \alpha T$ is the electron-electron scattering length and $K_{\text{micro}}\left( b, \gamma \right)$ is given by
\begin{align}\label{eq:problem:micro:Kmicro}
 K_{\text{micro}}&(b, \gamma ) = \nonumber \\
 = &\frac{\pi}{4} \int_{-\infty}^{\infty} \frac{dz}{\cosh^2(\pi z)} \int_0^\infty dx \int_0^\infty dy \times \nonumber \\
 & \times \mathcal{R} (x,z) \mathcal{R} (y,z) \mathcal{R} (x+y+xy,z) \times \nonumber \\
 & \times \Gamma_{WL} \left(b,\gamma L(x),\gamma L(y),\gamma ( L(x) + L(y) )\right).
\end{align}
Here $L(x) = \ln(1+x)$, $\gamma=\Delta k\, l_{ee}/2$ and
\begin{align}\label{eq:defR}
 \mathcal{R}(x,z) =&_2 F_1 ( 1/2 + i z, 1/2, 1; -x) \times \nonumber \\
 \times &_2 F_1 ( 1/2 - i z, 1/2, 1; -x),
\end{align}
where $_2F_1 (a,b,c;z) $ is the Gauss hypergeometric function.

The integral in Eq.~\eqref{eq:problem:micro:Kmicro} must be evaluated numerically.  In the next subsection, we will show this result, and compare it to the far simpler expression \eqref{eq:problem:phen:Kphen} calculated via the phenomenological approach.

\subsection{Comparison between phenomenological and microscopic calculations}
\label{sec:problem:comparison}

\begin{figure*}
 \begin{center}
 \subfigure[]{ \includegraphics[width=0.4\textwidth]{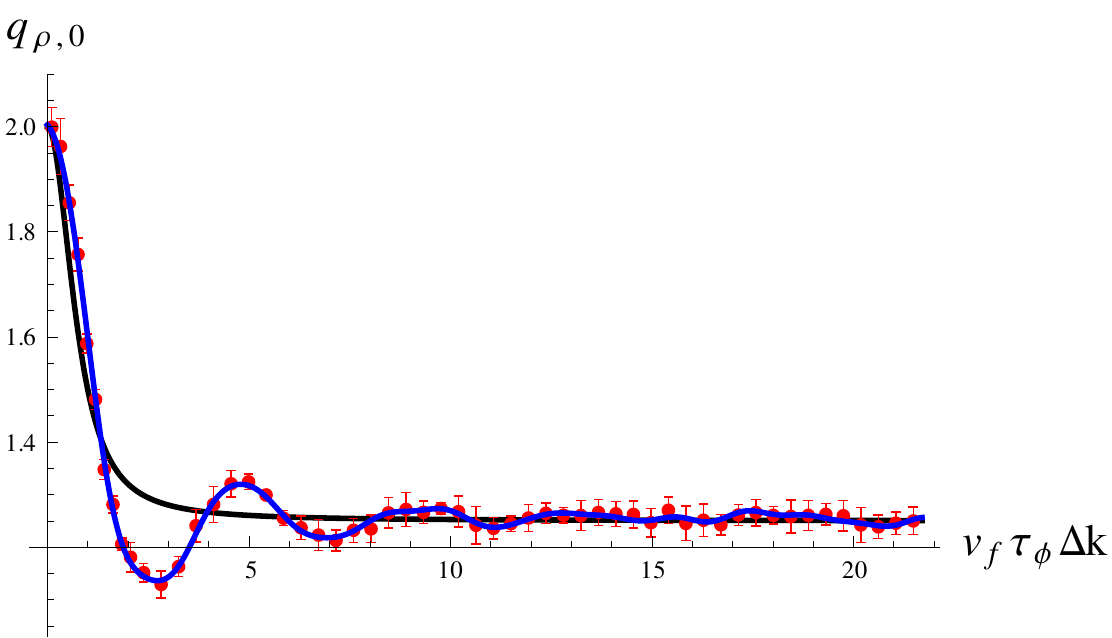}} \qquad \subfigure[]{ \includegraphics[width=0.4\textwidth]{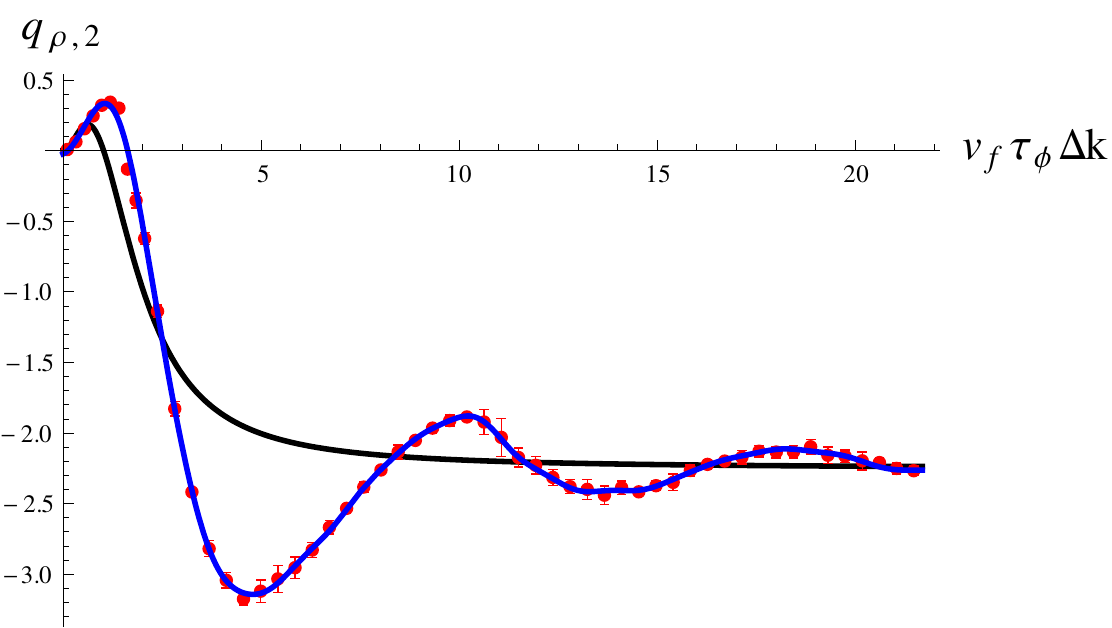}}\\
 \subfigure[]{ \includegraphics[width=0.4\textwidth]{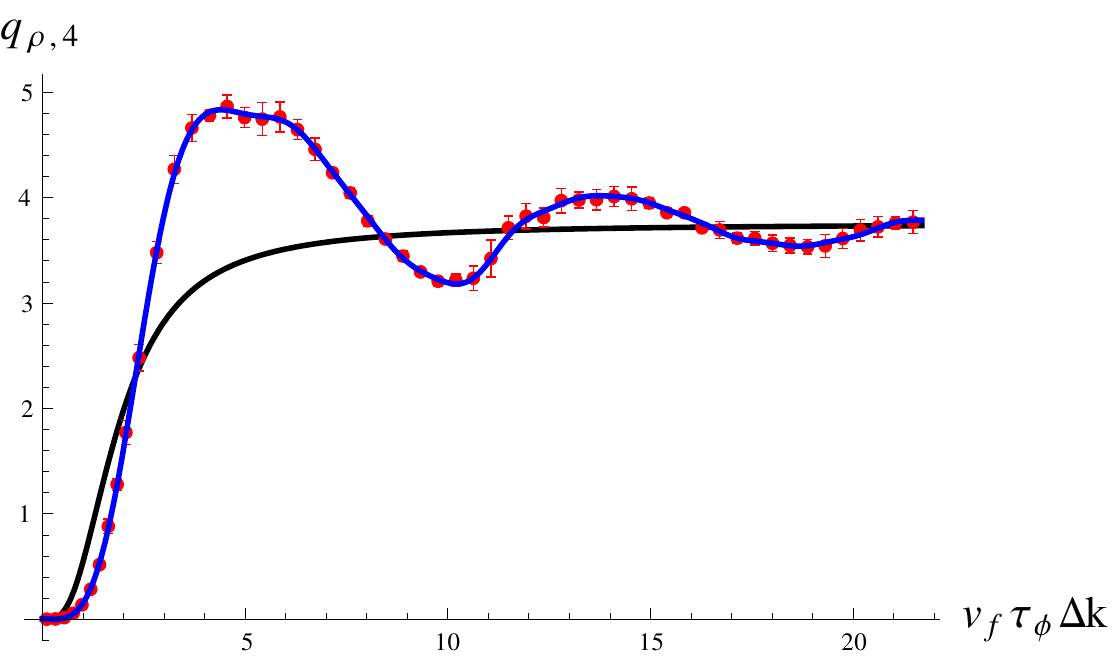}} \qquad \subfigure[]{ \includegraphics[width=0.4\textwidth]{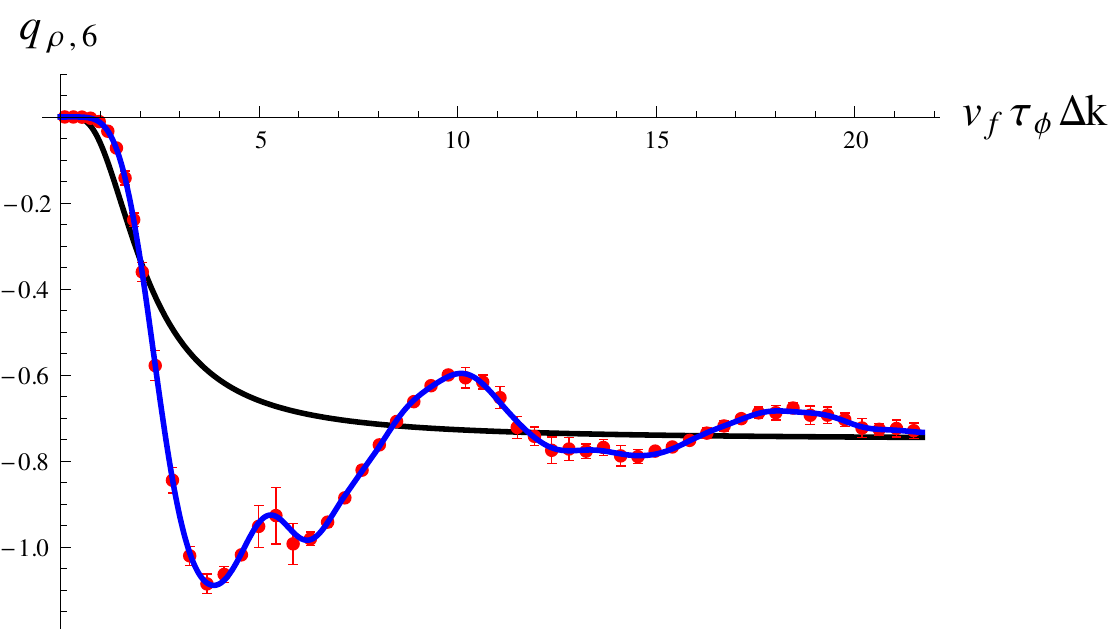}}
\end{center}
\caption{[Color online] Comparison between $q_{\text{phen},i} (v_F \tau_\phi \Delta k)$ (black solid line) and numerical values of $q_{\text{micro},i}(v_F \tau_\phi \Delta k)$ (red dots) as defined by Eq.~\eqref{eq:defqs}.  The blue curve is a spline fit to the numerical data points.  The microscopic $q's$ have been scaled so that $q_0=2$, in accordance with the phenomenological result.  This scaling also gives the relationship between dephasing time $\tau_\phi$ and electron scattering length $l_{ee}$ which fixes the horizontal scaling of the microscopic result (see main text).  Panel (a) after appropriate scaling also has physical meaning as the  weak localization correction to the conductivity of the ladder in the absence of an external magnetic field -- see section \ref{sec:discussion} of the main text.}
 \label{fig:problem:comparison:qPlots}
\end{figure*}

So far, we have calculated the weak localization correction of the spinless two-leg ladder in a phenomenological and a microscopic approach, the results are given in Eqs. \eqref{eq:problem:phen:WLcorrection} and \eqref{eq:problem:micro:WLcorrection_final}. Both results have a similar structure, even the functions $K_{\text{phen/micro}}\left( b, \gamma \right)$ have the same structure,
\begin{equation}\label{eq:defqs}
 K_{\rho} (b, \gamma) = q_{\rho,0}(\gamma) + b^2 q_{\rho,2}(\gamma) + b^4 q_{\rho,4}(\gamma) + b^6 q_{\rho,6}(\gamma),
\end{equation}
where $\rho = \text{phen} / \text{micro}$, $\gamma = v_F \tau_\phi \Delta k$ in the phenomenological and $\gamma = \Delta k l_{ee} /2$ in the microscopic case and the $q_{\rho,i}(\gamma)$ can be obtained from the different terms in Eqs. \eqref{eq:problem:phen:Kphen} and \eqref{eq:problem:micro:Kmicro}.  Note that while this may look like the first few terms in a power series expansion in (dimensionless) magnetic field $b$, the series naturally terminates due to our limitation to loops containing only three impurities.

The two-parameter functions $K_{\rho}(b,\gamma)$ may thus be expressed in terms of a linear combination of four functions of a single parameter, making comparison between the phenomenological and microscopic results easy.  The only thing to fix is the relationship between the dephasing time $\tau_\phi$ introduced by hand into the phenomenological model, and the electron-electron scattering length $l_{ee}$ that was a natural parameter in the microscopic approach.  In fact, this relationship may already be fixed by looking at the limiting value of $\Delta\sigma_{WL}$ when $\Delta k=0$, i.e. the limit of zero inter-chain hopping.  Fixing $\Delta\sigma_{WL}^\text{micro}(\Delta k =0) = \Delta\sigma_{WL}^\text{phen}(\Delta k =0)$ yields the relationship
\begin{equation}
\tau_\phi \simeq \: 1.09 \frac{l_{ee}}{v_F} = \frac{1.09}{\pi \alpha T}.\label{tauphiint}
\end{equation}

We can now plot the four functions functions $q_{\text{phen},i}(\gamma)$ and $q_{\text{micro},i}(\gamma)$ in Fig.~\ref{fig:problem:comparison:qPlots}.  Note that having established the relationship between $\tau_\phi$ and $l_{ee}$, this comparison has \textit{no fitting parameters} -- the numerical values obtained via Monte Carlo evaluations of the integrals \eqref{eq:problem:micro:Kmicro} are simply plotted alongside the analytic curves obtained from  \eqref{eq:problem:phen:Kphen}.  This comparison demonstrates two things -- firstly, the overall shape of the curves is very well captured by the phenomenological approach; and secondly, the microscopic calculation shows that these functions also contain an oscillating component not captured in the simple phenomenological calculation.  

The fact that these oscillations are not seen in the phenomenological picture means that they are related to features in the correlations of the two-leg ladder beyond dephasing, and nothing to do with normal magneto-resistance oscillations in two dimensions which originate from the Landau levels (in fact, the basically one-dimensional two-leg ladder does not exhibit Landau levels).   In Appendix \ref{sec:oscillations}, we demonstrate that the important feature giving rise to the oscillations is that of pseudospin-charge separation, i.e. the fact that the two leg ladder has two (collective) modes propagating with different velocities.  In the next section, we show how these oscillations translate into oscillations in the magnetoresistance, and give a more physical description of their origin.

\section{Discussion of weak localization and magneto-conductance}
\label{sec:discussion}

We turn now to an examination of the properties of the weak localization correction given by Eq.~\eqref{eq:problem:micro:WLcorrection_final},
\begin{equation}\label{eq:discussion:WLcorrection}
 \Delta \sigma_{WL} = -\frac{1}{8} \sigma^{(1)}_{D} \left( \frac{\tau_\phi}{\tau} \right)^2 \tilde{K} (\tau_\phi, t_\perp, f),
\end{equation}
where as usual, $\sigma^{(1)}_{D}$ is the Drude conductivity of a single chain and $\tilde{K} (\tau_\phi, t_\perp, f)$ is either the scaled microscopic result $\tilde{K} = (1.09)^{-2} K_\text{micro}$ or the phenomenological approximation to this $\tilde{K} = K_{\text{phen}}$.  These functions are given by Eqs. \eqref{eq:problem:phen:Kphen} and \eqref{eq:problem:micro:Kmicro}, where $b$ and $\Delta k$ are given by Eqs.~\eqref{eq:ladder:alpha} and \eqref{eq:ladder:Dk}; and for the microscopic case the relationship between $\tau_\phi$ and $l_{ee}$ is given in Eq.~\eqref{tauphiint}.  Ultimately, $\tilde{K}(\tau_\phi, t_\perp, f)$ depends only on the dimensionless ratios
\begin{equation}
\tilde{K}(\tau_\phi, t_\perp, f) = \tilde{K}\left( \tau_\phi t_\perp, \frac{v_F \pi f}{t_\perp}\right),
\end{equation}
however to keep the discussion more transparent, we retain the more physical original parameters.

The weak localization correction of the spinless Luttinger liquid is given by\cite{Gornyi-Mirlin-Polyakov-2007}
\begin{equation}
 \Delta \sigma_{WL}^{\text{LL}} = -\frac{1}{8} \sigma^{(1)}_{D} \left( \frac{\tau_\phi}{\tau} \right)^2,
\end{equation}
which differs from Eq.~\eqref{eq:discussion:WLcorrection} only by the function $\tilde{K}(\tau_\phi, t_\perp, f)$.  In order to investigate the properties of the weak localization correction, we examine the function  $\tilde{K}(\tau_\phi, t_\perp, f)$ and relate the results to the case of a single chain.  We discuss first the smooth part of the weak localization correction, for which we can use the analytic results from the phenomenological calculation $K_\text{phen}$, which matches the microscopic value in all limiting cases.  We then discuss the oscillations on top of this, for which we must switch to the results of the numerical integral, $K_\text{micro}$.

\subsection{Weak localization correction in the absence of a magnetic field}

In the absence of interchain hopping we find
\begin{equation}
 K_{\text{phen}} (\tau_\phi, t_\perp = 0, f) = 2,
\end{equation}
which means that the weak localization correction of the two-leg ladder is twice as large as the correction of the single chain. This is the result one would expect in the case of two decoupled chains.  Furthermore, this result is independent of magnetic field which must be so as for two decoupled chains, the magnetic field is a pure gauge.  However, this result is a non-trivial check on the calculation, as for the case $t_\perp=0$ in the band picture, zero magnetic field corresponds to $b=0,\;\Delta k=0$ while finite magnetic field corresponds to $b=1,\;\Delta k\rightarrow \infty$.  From Eq.~\eqref{eq:problem:phen:Kphen}, we then see that this gauge independence corresponds to the statement $5+1+2=5-9+15-3$.  A simple check confirms that the related equality is also satisfied by $K_\text{micro}$.

When interchain hopping is turned on but the magnetic field is absent, we find
\begin{align}\label{eq:discussion:limit_nomagnetic}
  K&_{\text{phen}}(\tau_\phi,t_\perp,f=0)  \nonumber \\
    &= \frac{1}{4} \left[ 5 + \frac{1}{1+(2t_\perp \tau_\phi)^2} + \frac{2}{(1+(2t_\perp \tau_\phi)^2)^2} \right],
\end{align}
which is smaller than 2 for non-zero $t_\perp \tau_\phi$. Hence, the weak localization correction of the two-leg ladder is smaller than the correction of two uncoupled chains. This result can be attributed to dimensionality, since localization effects become less pronounced with increasing dimensionality of the system.\cite{Abrahams-Anderson-Licciardello-Ramakrishnan-1979} In the limit of strong interchain hopping, $\tau_\phi t_\perp \gg 1$, we find $K_{\text{phen}}(\tau_\phi,t_\perp,f=0) \rightarrow 5/4$. The weak localization correction is still stronger than in the case of a single chain, but we cannot give a simple physical explanation of the factor $5/4$.

The weak localization correction in the absence of a magnetic field \eqref{eq:discussion:limit_nomagnetic} coincides with the function we previously called $q_0$, and is plotted in Fig.~\ref{fig:problem:comparison:qPlots}a as a function of $v_F\tau_\phi\Delta k = 2\tau_\phi t_\perp$ when $f=0$; which shows the smooth interpolation between the limits $2$ and $5/4$.  The figure also shows the equivalent result from the microscopic theory, which more or less follows the smooth phenomenological curve with some oscillations on top of this which die away as $\tau_\phi t_\perp \rightarrow \infty$.  One can understand this with the following simple picture: the parameter $\tau_\phi t_\perp$ is roughly speaking the number of times an electron can hop from one chain to the other before losing it's coherence.  When this number is small, the fact that the charge and pseudo-spin modes propagate at specific different velocities gives rise to coherent oscillations in the weak localization correction.  If this number is too large, the average over different disorder positions means such oscillations are washed out and only the smooth part (as captured by the phenomenological theory) remains.

\subsection{Magneto-conductivity -- smooth part}\label{sec:mono_nonmono}

\begin{figure}
\begin{center}
\includegraphics[width=0.3\textwidth]{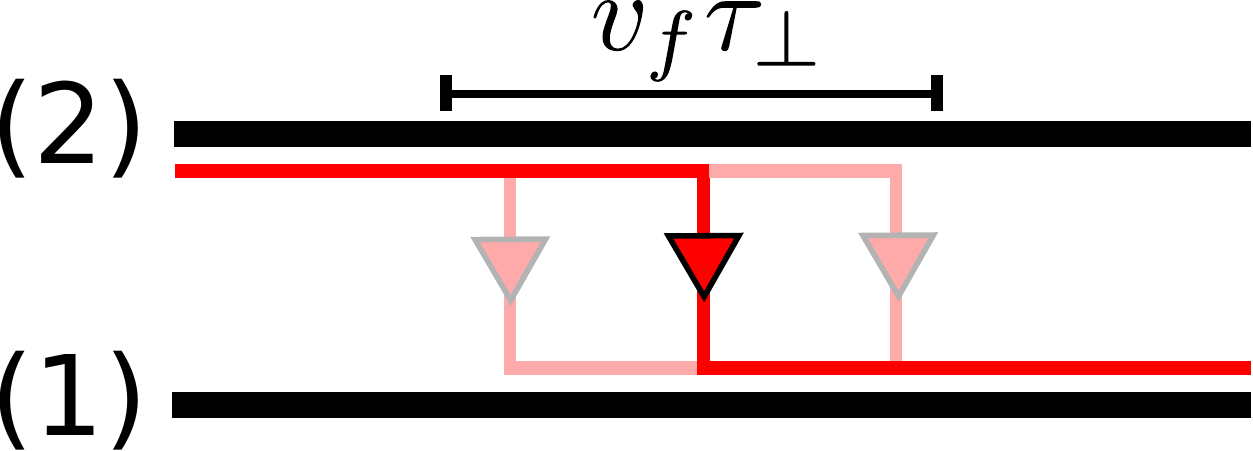}
\end{center}
\caption{Hopping from chain 2 to chain 1 in the continuous ladder model. Hopping can occur on a length scale of $l_{\perp}$, as shown by the shaded lines.}
\label{fig:discussion:ladder_hopping}
\end{figure}

We first consider the case of a strong magnetic field, where we find
\begin{equation}
 K_{\text{phen}} (\tau_\phi, t_\perp , f \rightarrow \infty) \rightarrow 2,
\end{equation}
which is the same as the limit of two uncoupled chains. Surprisingly, the weak localization correction is stronger in the presence of a magnetic field than without it, so the magnetic field \textit{enhances} localization.

This rather counterintuitive result is a consequence of the specific geometric setup of the two-leg ladder, and can actually be understood rather easily, in a similar vein to the dying away of oscillations discussed above.  A single electron propagating along one of the chains can hop to the other one on a typical length scale of $l_{\perp} = v_F \tau_\perp$, where we have introduced the average hopping time $\tau_\perp = t_\perp^{-1}$ (see Fig.~\ref{fig:discussion:ladder_hopping}). Since $t_\perp$ acquires a position-dependent Peierls phase in the presence of a magnetic field, the hopping parameter feels a different phase, depending where the hopping takes place. These hopping events interfere destructively and lead to an effective suppression of interchain hopping, and hence is equivalent to the limit of two decoupled chains. This behavior is already seen in Eq.~\eqref{eq:ladder:chain-band-mapping}, where the inter-chain Green functions vanish for large magnetic fields, $b \rightarrow 1$.

\begin{figure*}
 \begin{center}
 \subfigure[]{ \includegraphics[width=0.32\textwidth]{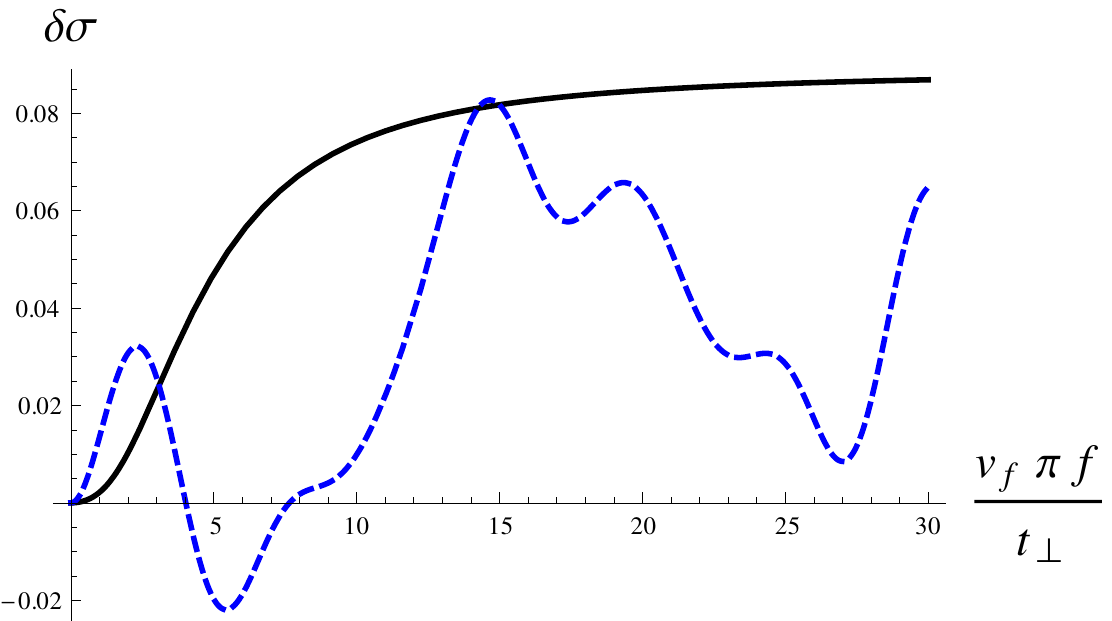}}
 \subfigure[]{ \includegraphics[width=0.32\textwidth]{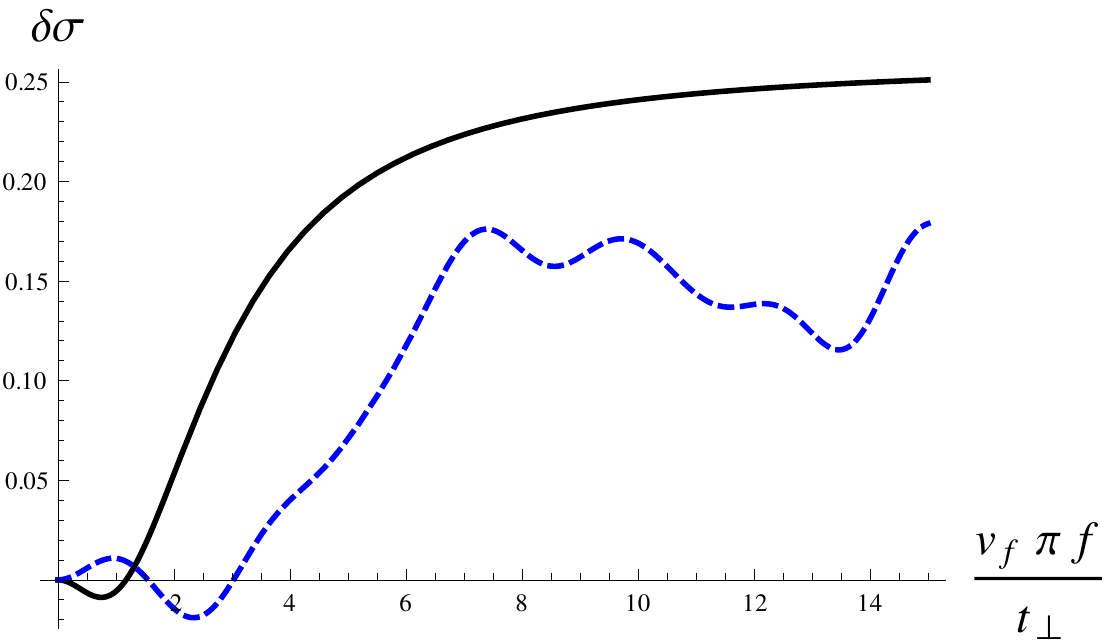}}
  \subfigure[]{ \includegraphics[width=0.32\textwidth]{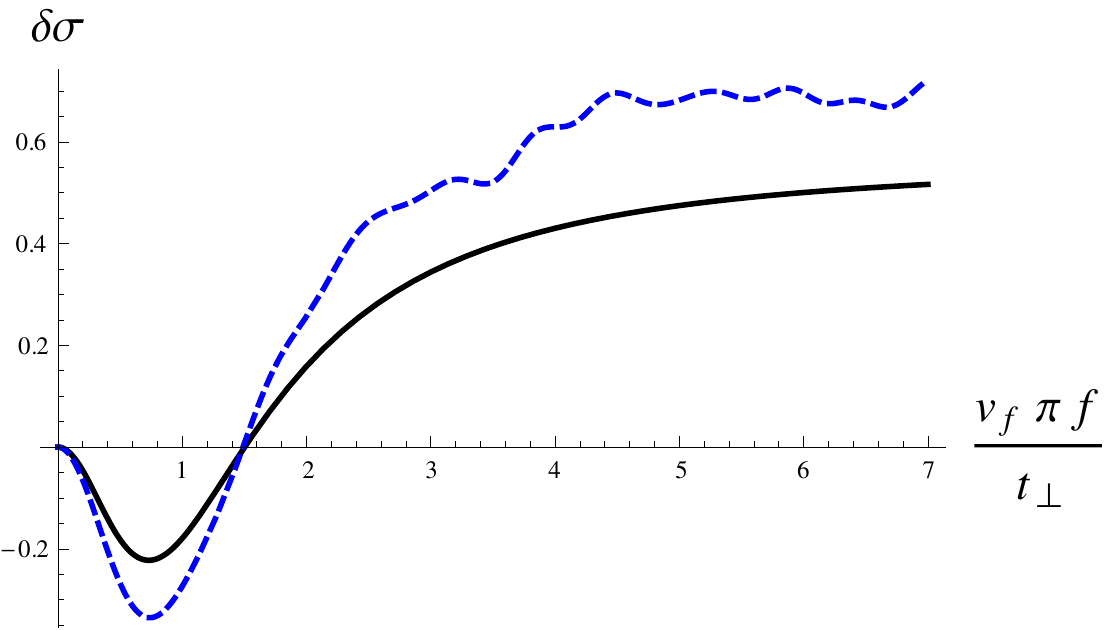}}
\end{center}
 \caption{[Color online] Plots of dimensionless magneto-conductivity $\delta \sigma (\tau_\phi, t_\perp, f)$ against the generalized magnetic field, $v_F \pi f / t_\perp$ for (a) $t_\perp \tau_\phi = 0.2$, (b) $t_\perp \tau_\phi = 0.4$ and (c) $t_\perp \tau_\phi = 1.5$.  The black solid lines are obtained from the phenomenological calculation, while the blue dashed lines correspond to the equivalent microscopic results, approximating the numerical functions by the spline fits as obtained in Fig.~\ref{fig:problem:comparison:qPlots}.  The black curves show the evolution from a monotonic to non-monotonic magneto-conductivity as $t_\perp \tau_\phi$ is increased, as discussed in Sec.~\ref{sec:mono_nonmono}.  The blue curves show the magneto-conductivity oscillations on top of this, as discussed in Sec.~\ref{sec:mcoscillations}.}
 \label{fig:discussion:dsigma_plots}
\end{figure*}

We now define the dimensionless magneto-conductivity
\begin{equation}
 \delta \sigma (\tau_\phi, t_\perp, f) = \frac{\Delta \sigma_{WL}(\tau_\phi, t_\perp, f) - \Delta \sigma_{WL}(\tau_\phi, t_\perp, 0)}{\Delta \sigma_{WL}(\tau_\phi, t_\perp, 0)}
\end{equation}
to study the behaviour of the weak localization correction for finite magnetic fields at a given $\tau_\phi t_\perp$.  We have plotted $\delta \sigma (\tau_\phi, t_\perp, f)$ for different values of $t_\perp \tau_\phi$ in Fig. \ref{fig:discussion:dsigma_plots}. 

For small values of $t_\perp \tau_\phi$, the limit of two decoupled chains is achieved in a monotonic way. For large values of $t_\perp \tau_\phi$, however, the behaviour is non-monotonic.  This comes about from the competition of the above peculiar effect of magnetic-field enhanced localization, and the more usual decohering effect of the magnetic field that reduces quantum corrections to conductivity in higher dimensions.\cite{Abrahams-Anderson-Licciardello-Ramakrishnan-1979}.  When $t_\perp \tau_\phi$ is large, the dominant contribution to the weak localization correction comes from loops in which the electron hops coherently many times between the two legs of the ladder -- and hence for small fields such loops will feel the normal decoherence effect of a flux penetrating such loops.  Only at larger magnetic fields does the peculiar enhancement of weak localization in this geometry take over through the effective decoupling of the chains.

A numerical analysis shows that the transition from monotonic to non-monotonic behaviour occurs in the phenomenological model at $t_\perp \tau_\phi \simeq 0.266$.  However, Fig.~\ref{fig:discussion:dsigma_plots} clearly also shows that at these small values of $t_\perp \tau_\phi$ the oscillations in magneto-conductivity arising from correlation effects are large; so the neat change in behavior predicted by the phenomenological calculation is not possible to see.  The oscillations will be discussed below.

\subsection{Magneto-conductivity -- oscillating part}\label{sec:mcoscillations}

We previously showed in Fig.~\ref{fig:problem:comparison:qPlots} that the microscopic calculation yields oscillations in the weak localization correction which were not seen in the simple phenomenological calculation -- thereby demonstrating that these oscillations go beyond the single particle (with dephasing) picture, and depend in a more essential way on the nature of the strong correlations ubiquitous in one-dimensional interacting systems.  In Appendix \ref{sec:oscillations}, we showed on a technical level that these oscillations have their origin in the pseudospin-charge separation; while in Fig.~\ref{fig:discussion:dsigma_plots} we saw how this translates into oscillations in the magneto-conductivity which are particularly pronounced when $t_\perp \tau_\phi\lesssim 1$.

In fact, there have been a number of predictions of the effects of spin-charge separation on transport in clean Luttinger liquids.  Some examples include time-resolved measurements of the propagation of pulses,\cite{Ulbricht-Schmitteckert-2009} or interference effects in finite rings\cite{Jagla-Balseiro-1993,Friederich-Meden-2008,Rincon-Aligia-Hallberg-2009} with an Aharonov-Bohm flux.  In fact,  for sufficiently short $t_\perp \tau_\phi$, the weak localization in our ladder system can be thought of as qualitatively similar to a finite ring as in this limit one expects simple loops with only two inter-chain hopping events before the electrons become dephased.  Therefore, the oscillations we predict in magneto-conductivity have essentially the same physical origin as the manifestation of spin-charge separation in Aharonov-Bohm rings.\cite{Jagla-Balseiro-1993,Friederich-Meden-2008,Rincon-Aligia-Hallberg-2009}  It is an intriguing result that physically observable effects of the interaction-induced correlations in the two leg-ladder system appear more naturally in transport in the disordered case than in the clean case.

However, as $t_\perp \tau_\phi$ becomes larger, many more coherent inter-chain hops of the electron take place, and the beautiful interference effects arising from two different velocities become washed out as previously explained.

\subsection{Experimentally measurable conductivity}

In an experimental setup, the parameters easily varied are temperature and magnetic field, while one measures the full conductivity including the Drude component
\begin{equation}
\sigma(T,B) = \sigma_D(T) + \Delta \sigma_{WL}(T,B).
\end{equation}
We remember that the weak temperature dependence of the Drude conductivity $\sigma_D$ comes from the Luttinger liquid renormalization of the transport scattering rate, while the strongest temperature dependence of the WL correction is through the temperature dependence of the dephasing time $\tau_\phi\propto 1/T$, which not only enters the functional dependence on magnetic field but appears as a prefactor in the strength of the quantum correction.  In a typical single experiment, other material constants such as interaction strength, or inter-chain hopping, will remain constant.

\begin{figure}
 \begin{center}
\includegraphics[width=0.4\textwidth]{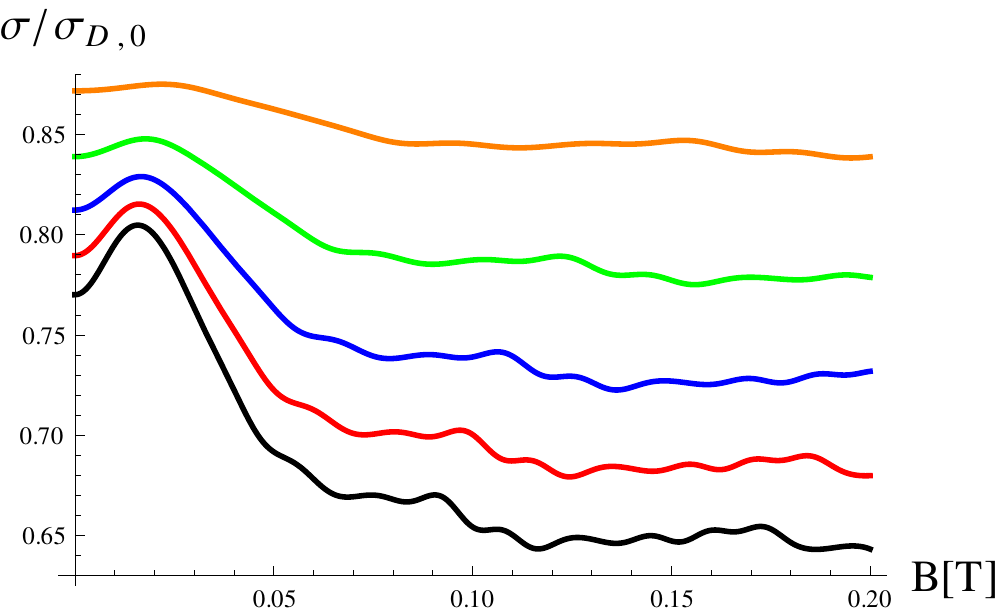}
\end{center}
 \caption{[Color online] Plot of full conductivity (including Drude component) as a function of magnetic field for various temperatures, from top to bottom $T=50,40,35,32,30K$.  The other parameters used in the plot are $\alpha=0.01$, $\Lambda=10000K$, $v_F=10^8$cm/s, $a_y=1$nm, $t_\perp=1K$ and $1/\tau^{(0)}_\mathrm{tr}=1K$.  The normalization of the vertical axis $\sigma_{D,0}$ is the Drude conductivity in the high temperature limit.}
 \label{fig:discussion:fullconductivity}
\end{figure}

In order to relate to potential real experiments, we plot in Fig.~\ref{fig:discussion:fullconductivity} the conductivity against magnetic field for various temperatures, being careful to choose a parameter range which is both physically reasonable, and in which the approximations in our calculations are valid -- the exact parameters chosen are shown in the caption to the figure.  As expected, the weak localization corrections (and hence the magnetic field dependence) become stronger as temperature is reduced.  Furthermore, we see clearly the fact that, as a general trend, conductivity decreases with increasing magnetic field -- the effect we call magnetic field enhanced localization.  We also see oscillations in conductivity as a function of magnetic field -- which we remember are not due to conventional effects such as mesoscopic fluctuations (we average over disorder in our calculations) or Landau levels (which do not exist in the two-leg ladder); but in fact have their origin in correlations, namely pseduo-spin charge separation.

It is instructive to compare our results to the experiment on nanotubes by Man and Morpurgo.\cite{nanotube_magnetoresistance}  One should not compare details too closely as the geometric set-up of a nano-tube differs from that of a two-leg ladder; however it is already striking how similar some of the features of our Fig.~\ref{fig:discussion:fullconductivity} and the experimental curve (Fig. 3 of Ref.~\onlinecite{nanotube_magnetoresistance}) look.  It remains work for the future to adapt this calculation to the nanotube geometry relevant for this particular experiment; and to see whether this type of orbital magneto-resistance is applicable to single wall carbon nanotubes.

\section{Summary}

In summary, we have calculated the weak localization correction to charge transport and the magneto-resistance for a model system of  a disordered two-leg ladder.  We have limited ourselves to the high-temperature regime defined by the two conditions:
\begin{enumerate}
\item Temperature $T$ should be greater than any interaction induced gaps.
\item Dephasing time $\tau_\phi \propto (\alpha T)^{-1}$ should be much less than the mean free time $\tau_\mathrm{tr}$ of diffusion in the (weakly) disordered system.
\end{enumerate}

In this regime, we find that the weak localization correction is a universal function of $t_\perp \tau_\phi$ and $f/t_\perp$.  Our results can be summarized as follows:
\begin{enumerate}
\item In the absence of a magnetic flux, the weak localization correction to the Drude conductivity is a decreasing function of the size of the inter-chain hopping $t_\perp \tau_\phi$ from $2$ at $t_\perp=0$ to $5/4$ at $t_\perp\rightarrow \infty$ in units of the single chain correction.  This crossover is simply an aspect of the phenomena that as dimension is increased (in this case going from one pure one-dimensional chains to two coupled chains), weak localization corrections decrease -- this is plotted in Fig.~\ref{fig:problem:comparison:qPlots}a.
\item There are two main competing effects contributing to magnetoresistance.  The first is the decoherence of weak localization loops by the magnetic field, leading to diminishing weak localization correction as flux is increased.  The other effect is the tendency of the magnetic field to decouple the two chains, and thus enhancing the weak localization correction.  For $t_\perp \tau_\phi$ not too small, the two effects compete and lead to a non-monotonic magnetoresistance -- this is shown in Fig.~\ref{fig:discussion:dsigma_plots}.
\item When $t_\perp \tau_\phi$ is small, there is a third observable effect seen in the magnetoresistance, which is coherent oscillations due to the different velocities of charge and pseudo-spin excitations in the two-leg ladder -- this is also shown in Fig.~\ref{fig:discussion:dsigma_plots}.
\end{enumerate}
Finally, Fig.~\ref{fig:discussion:fullconductivity} puts all of these results together, and shows a typical example of what we predict would be seen in an experimental measurement on such a system.

We expect that these results, though derived for a specific toy model, are more general than this as we can identify the basic physical origin of each effect.  In particular, results 1 and 2 are even seen in a simple phenomenological model where the important features are a phenomenological dephasing time and the system geometry.  Our microscopic calculation backed up these general observations, as well as showing the beautiful additional effect of magnetoresistance oscillations arising from (pseudo)spin-charge separation, which we believe goes beyond the specific geometry of the two-leg ladder and may even be seen in e.g. nanotubes.

We acknowledge useful discussions with A.~Yashenkin and M.~Sch\"utt, and support from the German-Israeli foundation.  IVG acknowledges illuminating discussions with D.~Polyakov during the preparation of Ref.~\onlinecite{Yashenkin-Gornyi-Mirlin-Polyakov-2008} about a related ladder model that turns out not to exhibit orbital magnetoresistance.  STC acknowledges early discourses with N.~Hussey and A.~Narduzzo which in part motivated the present work.  MPS acknowledges useful discussions with T.~Sproll.


\appendix

\section{Functional bosonization}
\label{sec:appendix:funcbos}
In this appendix we present the method of functional bosonization for a disordered spinful Luttinger liquid. The method has been used in the spinless\cite{Gornyi-Mirlin-Polyakov-2005,Gornyi-Mirlin-Polyakov-2007} and spinful\cite{Yashenkin-Gornyi-Mirlin-Polyakov-2008} case; a more detailed derivation is presented in the given references. 

\subsection{Basics}

The key ingredient of functional bosonization is a Hubbard-Stratonovich transformation which decouples the four-fermion interaction by introducing a bosonic field $\varphi(x,\tau)$. This field has the correlation function
\begin{equation}
 \left< \varphi(x,\tau) \varphi(0,0) \right> = V(x,\tau),
\end{equation}
where $V(x,\tau)$ is the dynamically screened interaction. This leads to a quadratic action for the Fermions; integrating them out then gives the usual bosonized action -- however this suffers from the disadvantage that fermionic operators are non-linear (and in some representations even non-local) in terms of the bosons, which creates difficulties in creating a diagrammatic expansion for disorder.  In contrast, functional bosonization retains both the fermionic and bosonic degrees of freedom in the action; we will now outline how this overcomes the above difficulty and one can treat further perturbations to the system such as disorder.

In one dimension, fermionic and bosonic fields can be decoupled in the action by the gauge transformation
\begin{equation}\label{eq:appendix:funcbos:psi_gauge}
 \psi_{\mu \sigma} (x,\tau) = \psi_{\mu \sigma} (x,\tau) \exp \left[ i \theta_{\mu}(x,\tau) \right],
\end{equation}
where $\mu=R/L$ is the chirality, $\sigma$ is the (pseudo-)spin and the phase $\theta_{\mu}(x,\tau)$ obeys
\begin{equation}
 \left( \partial_\tau - i\mu v_F \partial_x \right) \theta_{\mu}(x,\tau) = \varphi_{\mu}(x,\tau).
\end{equation}
\noindent

When we write down the full Green functions in terms of the fermionic operators given in Eq.~\eqref{eq:appendix:funcbos:psi_gauge} and perform the Gaussian averages over the $\theta(x,\tau)$-fields, we find [see e.g. Ref.~\onlinecite{Yurkevich-2001}]
\begin{equation}\label{eq:geebee}
 G_{\mu}(x,\tau) = g_{\mu}(x,\tau) \exp \left[ - B_{\mu \mu} (x,\tau) \right].
\end{equation}
\noindent
Here, the free Green function (ignoring for now phase factors $e^{\pm i k_F x}$ -- see main text) are given by
\begin{equation}\label{eq:freegee}
 g_{\mu} (x,\tau) = -\mu \frac{i T}{2 v_F} \frac{1}{\sinh \left[ \pi T (x/v_F + i \mu \tau) \right]}
\end{equation}
while the bosonic correlator $B_{\mu \nu} (x,\tau)$ defined as
\begin{equation}\label{eq:appendix:funcbos:B_generic_realspace}
 B_{\mu \nu} (x,\tau) = \left< \left[ \theta_{\mu} (0,0) - \theta_{\mu} (x,\tau) \right] \theta_{\nu} (0,0) \right>.
\end{equation}
is given by
\begin{align}\label{eq:appendix:funcbos:B_generic_momentumspace}
 B_{R,R/L} (x,\tau) &= T \sum_{n} \int \frac{dq}{2\pi} \left( e^{iqx - i\Omega_n \tau} - 1 \right) \times \nonumber \\ 
		    &\times \frac{ V (q,i\Omega_n) }{(v_F q - i \Omega_n)(\pm v_F q - i \Omega_n)}.
\end{align}
The remaining chiral combinations are given by
\begin{equation}
 B_{LL}(x,\tau) = B_{RR}(-x,\tau), \quad B_{LR}(x,\tau)=B_{RL}(x,\tau).
\end{equation}
In this expression, $\Omega_n = 2\pi n T$ is the bosonic Matsubara frequency, and the dynamically screened interaction $V(q,i\Omega_n)$ is given by
\begin{equation}\label{eq:appendix:funcbos:Vscreened}
 V(q,i\Omega_n) = g \frac{v_F^2 q^2 + \Omega_n^2}{u^2 q^2 + \Omega_n^2},
\end{equation}
where the charge velocity $u$ is given by
\begin{equation}
 u = v_F \sqrt{1 + 2g/\pi v_F}.
\end{equation}
Within our original model with a single interaction parameter, $g$, the (pseudo-)spin velocity remains unrenormalized and so is equal to $v_F$.

Substituting Eq.~\eqref{eq:appendix:funcbos:Vscreened} into Eq.~\eqref{eq:appendix:funcbos:B_generic_momentumspace} and executing the integral, one finds
\begin{align}\label{eq:bee}
 B_{RR}(x,\tau) &= -\frac{1}{2} \ln \eta (x,\tau) - \frac{\alpha_b}{4} \ln \varsigma (x,\tau), \nonumber \\
 B_{RL}(x,\tau) &= -\frac{\alpha_r}{4} \ln \varsigma (x,\tau),
\end{align}
where
\begin{gather}
 \varsigma(x,\tau) = \frac{(\pi T/\Lambda)^2}{\sinh \left[ \pi T (x/u+i\tau) \right] \sinh \left[ \pi T (x/u-i\tau) \right] }, \nonumber \\
 \eta (x,\tau) = \frac{v_F}{u} \frac{\sinh \left[ \pi T (x/v_F+i\tau) \right]}{\sinh \left[ \pi T (x/u+i\tau) \right]}. \label{eq:weee}
\end{gather}
The prefactors $\alpha_{b,r}$ in Eq.~\eqref{eq:bee} are given by
\begin{equation} 
\alpha_b = \frac{(u-v_F)^2}{2 u v_F} \approx \frac{\alpha^2}{2}, \qquad \alpha_r = \frac{u^2-v_F^2}{2uv_F} \approx \alpha,
\end{equation}
where $\alpha=g/\pi v_F$ is the dimensionless interaction strength introduced in Sec.~\ref{sec:interactions}.
In the limit of weak interactions, the prefactors follow the hierarchy
\begin{equation}
 \alpha_b \ll \alpha_r \ll 1,
\end{equation}
since $\alpha_b$ is quadratic in the interaction strength, whereas $\alpha_r$ is linear.

\begin{figure}
\begin{center}
\includegraphics[width=0.45\textwidth]{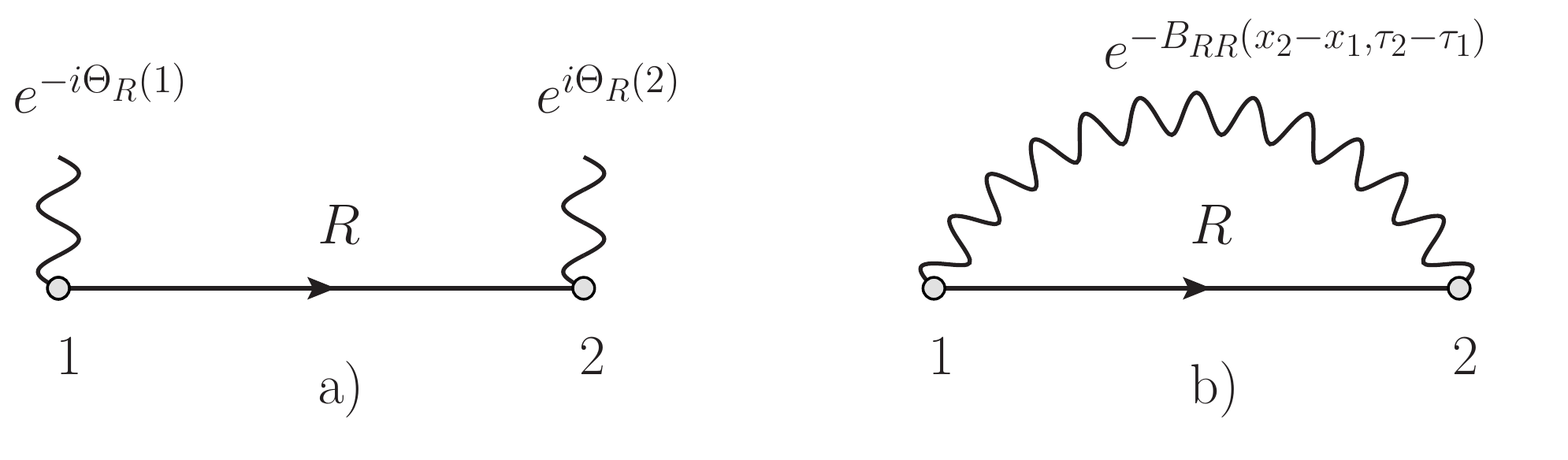}
\end{center}
\caption{Green function of a right-mover propagating from space-time point $1=(x_1,\tau_1)$ to $2=(x_2,\tau_2)$ before (a) and after (b) averaging over the $\theta(x,\tau)$-fields. The solid line is the free Green function, the wavy line stands for the factors $\exp [ -i \theta_R(1) ]$ and $\exp [ i \theta_R(2) ]$. After averaging over the $\theta$-fields, the points $1$ and $2$ become connected, the wavy line in (b) stands for the factor $e^{-B_{RR}(x_2-x_1, \tau_2 - \tau_1)}$.}
\label{fig:appendixFB:GFdiagram}
\end{figure}

\subsection{Diagrammatic expansion}\label{sec:diagrams}

\begin{figure}
\begin{center}
\subfigure[]{\includegraphics[width=0.45\textwidth]{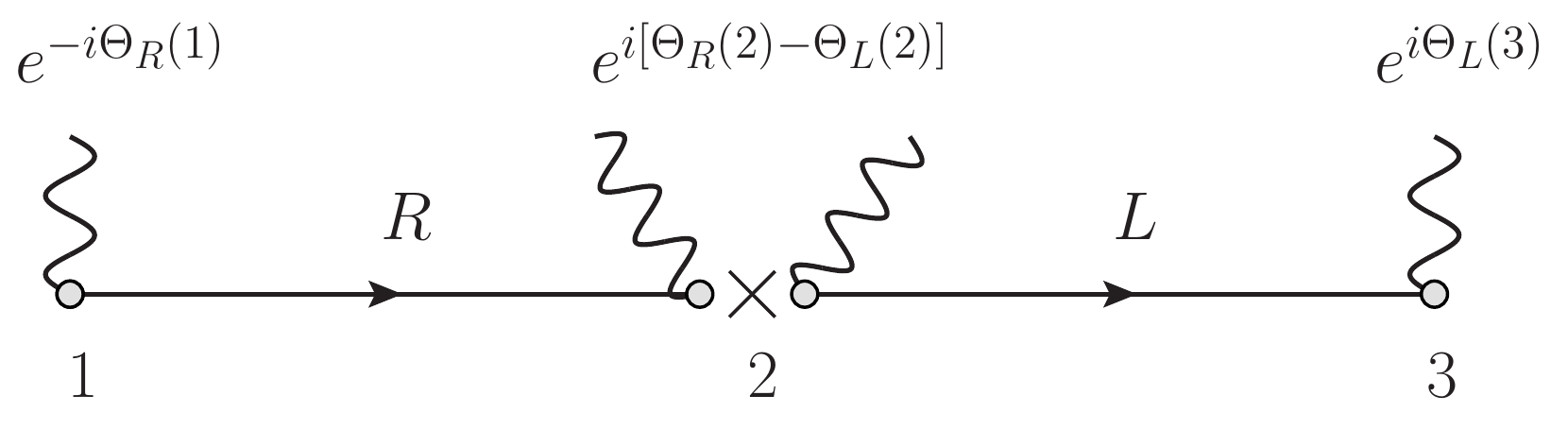}}
\subfigure[]{\includegraphics[width=0.30\textwidth]{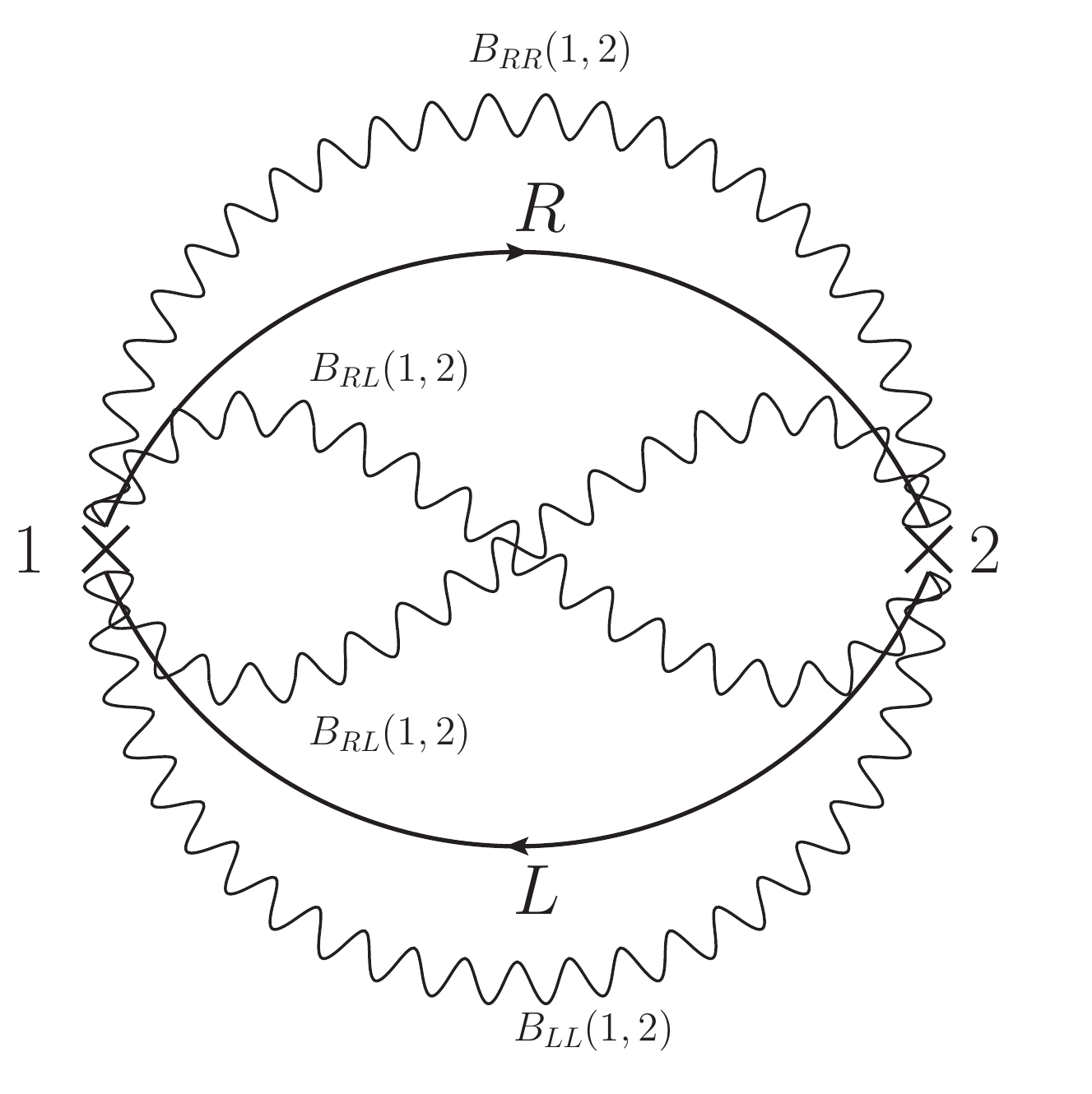}}
\end{center}
\caption{(a) Backscattering process of a right-mover on an impurity (denotes by a cross), both Green functions are dressed by $\theta$-fields. (b) The possible connections of the $\theta$-fields for averaging in the simplest loop; which yields a $Q$ factor, Eq.~\eqref{eq:defQ}, associated with the pair of vertices.  In a more complicated loop, each pair of vertices gives a factor $Q$ or $Q^{-1}$ (see main text).}
\label{fig:appendixFB:disorder_unconnect}
\end{figure}

Equation \eqref{eq:appendix:funcbos:psi_gauge} suggests the following diagrammatic technique: The $\theta(x,\tau)$-fields get attached to the starting and ending points of the free propagator, drawn as wavy lines in Fig.~\ref{fig:appendixFB:GFdiagram}a. After averaging over the $\theta$-fields, the two lines become connected and represent the correlator $e^{-B_{\mu\mu}}$ -- see Fig.~\ref{fig:appendixFB:GFdiagram}b. This diagrammatic technique can be extended to scattering off impurities in a straightforward way. A backscattering process is shown in Fig.~\ref{fig:appendixFB:disorder_unconnect}a, interaction is accounted by the $\theta$-fields attached to the free Green function. Averaging results in connecting all wavy lines pairwise in every different way.

In closed fermionic loops, the $B_{\mu \nu}$-correlators appear only in the combination (see Fig.~\ref{fig:appendixFB:disorder_unconnect})
\begin{equation}
 M(x,\tau) = B_{RR} (x,\tau) + B_{LL} (x,\tau) - 2 B_{RL} (x,\tau).
\end{equation}
As a consequence, each pair of backscattering vertices at space-time points $N$ and $N'$ contributes a factor of
\begin{equation}\label{eq:defQ}
 Q(x_N - x_{N'}, \tau_N - \tau_{N'}) = \exp \left[ M (x_N - x_{N'}, \tau_N - \tau_{N'}) \right]
\end{equation}
if the chiralities of the incident electrons at the vertices are the same, and a factor of $Q^{-1}$ if they are different.

\subsection{Green function in $(x,\epsilon)$ representation}\label{sec:GFs}

Combing Eqs.~\eqref{eq:geebee}, \eqref{eq:freegee}, \eqref{eq:bee} and \eqref{eq:weee}, and assuming $\alpha_b\ll 1$ (weakly interacting limit) so that the $\varsigma$ term may be neglected, the interacting Green function takes the form
\begin{gather}\label{eq:GFs1}
G_\mu(x,\tau) \approx -\mu \frac{i}{2\pi\sqrt{u v_F}} \\ \nonumber
\times \left\{ \frac{\pi T}{\sinh [\pi T ( x/v_F + i\mu \tau) ]} \frac{\pi T}{ \sinh [\pi T ( x/u + i\mu \tau) ]} \right\}^{1/2}.
\end{gather}

Within this approximation, the only singularity of $G_\mu$ as considered as a function of $\tau$ is a branch cut between $\tau=|x|/u$ and $|x|/v_F$ in the upper of lower-half plane depending on the sign of $\mu x$.  Choosing carefully the contour\cite{Yashenkin-Gornyi-Mirlin-Polyakov-2008} one can therefore represent the Fourier transform of the Green function with respect to $\tau$ as
\begin{equation}
\int_0^{1/T} d\tau e^{i\epsilon_n \tau} G_\mu(x,\tau) = G_\mu^r(x,i\epsilon_n) - G_\mu^a(x,i\epsilon_n),
\end{equation}
where $\epsilon_n=2\pi(n+\frac{1}{2})T$ is the fermionic Matsubara frequency and
\begin{eqnarray}
G_\mu^r(x,i\epsilon_n) &=& \theta(\epsilon_n) \theta(\mu x) {\cal G}(|x|,|\epsilon_n|) \nonumber \\
G_\mu^a(x,i\epsilon_n) &=& \theta(-\epsilon_n) \theta(-\mu x) {\cal G}(|x|,|\epsilon_n|). \label{eq:defGar}
\end{eqnarray}
Executing the integral gives ${\cal G}(x>0,\epsilon_n>0)$ as \cite{Yashenkin-Gornyi-Mirlin-Polyakov-2008}
\begin{equation}
{\cal G}(x,\epsilon_n) = \frac{e^{-\epsilon_n x/u + x/2l_{ee}}}{i\sqrt{u v_F}} {}_2F_1\left[ \frac{1}{2} + \frac{\epsilon_n}{2\pi T}, \frac{1}{2}, 1; 1-e^{2x/l_{ee}}\right]
\end{equation}
where ${}_2F_1(a,b,c;z)$ is the hypergeometric function and $l_{ee}=v_F/\alpha T$ is the electron-electron scattering length.  

For a full discussion of the properties of this Green function, see Ref.~\onlinecite{Yashenkin-Gornyi-Mirlin-Polyakov-2008}.


\section{Oscillations arising from different spin and charge velocities}\label{sec:oscillations}

In this appendix, we show mathematically the relationship between the phenomenological calculation Eq.~\eqref{eq:problem:phen:WLcorrection} and the microscopic result Eq.~\eqref{eq:problem:micro:WLcorrection_final}.  By making this comparison, we isolate the source of the oscillations in the weak localization correction as seen in the microscopic calculation, and show that these may be attributed to pseudo-spin-charge separation.

Starting from the integral for the microscopic calculation, Eq.~\eqref{eq:problem:micro:Kmicro}, we see that the $1/\cosh^2(\pi z)$ factor means that the dominant contribution comes around $z\approx 0$.  Therefore using the expansion of the hypergeometric function
\begin{equation}
{}_2F_1(1/2,1/2,1;-x) \approx \frac{1}{\pi\,x^{1/2}}\ln(16x) + o(x^{-3/2})\label{eq:hypergeometric_expansion}
\end{equation}
and ignoring the logarithmic corrections to the power law, we see that this integral may be approximated as\cite{Yashenkin-Gornyi-Mirlin-Polyakov-2008}
\begin{align}\label{eq:micro_phen}
 \frac{\Delta \sigma_{WL} }{\sigma^{(1)}_{D}} &\sim - \int_{0}^{\infty}\frac{dx_{a}}{l} e^{ - \frac{2 x_a}{l_{ee}} } 
 \int_{0}^{\infty}\frac{dx_{b}}{l} e^{- \frac{2 x_b}{l_{ee}}} \times \nonumber \\
					&\hphantom{\sim} \times \Gamma_{WL}(\Delta k x_a,\Delta k x_b,\Delta k (x_a+x_b)) \nonumber \\
					&\sim \left( \frac{l_{ee}}{l} \right)^2 K_{\text{phen}} (b, \Delta k l_{ee} ),
\end{align}
where we have returned to the original variables $x_a$ and $x_b$ which are physically distances between impurities.  We may now consider what will happen when the logarithmic factors present in Eq.~\eqref{eq:hypergeometric_expansion} are taken into account.  While these factors will change the true asymptotic behavior of the Green function, this is actually irrelevant for evaluating the integral above as the contribution from these regions is exponentially small.  We will shortly show that the slowly varying logarithm at intermediate length scales can actually be very well approximated by a slight modification of the power law in Eq.~\eqref{eq:hypergeometric_expansion}; which translates in the integral above to a slight modification of $l_{ee}$.  In fact, fitting this power law at this intermediate length scale (see the inset to Fig.~\ref{fig:feature}) gives us exactly the modification in Eq.~\eqref{eq:micro_phen} of
\begin{equation}
l_{ee}\rightarrow l_\phi = v_F \tau_\phi \approx1.09 l_{ee},
\end{equation} 
in full agreement with Eq.~\eqref{tauphiint} in the main text.  In this way, we have rederived the phenomenological result out of the microscopic model.  It is worth noting that this single number governs both the vertical and horizontal scaling in all four graphs in Fig.~\ref{fig:problem:comparison:qPlots}.

This may be seen even clearer by noticing that the above expansion essentially corresponds to writing the Green function of the electron as
\begin{equation}
G(x_a>0,\epsilon_n\approx 0) \rightarrow e^{-x_a/2l_{\phi}},
\end{equation}
which is exactly the form used for the phenomenological Green function, Eq.~\eqref{eq:problem:phen:GF-dephasing}.
\begin{figure}
\begin{center}
\includegraphics[width=0.35\textwidth]{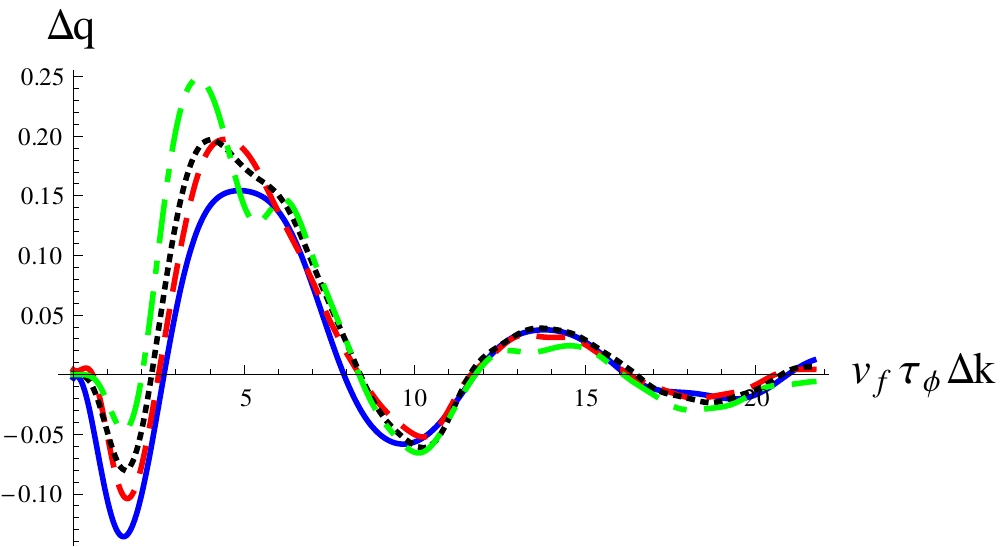}
\end{center}
\caption{[Color online] The differences between the microscopic and phenomenological results $\Delta q_i$ as a function of $\gamma=v_F \tau_\phi \Delta k$.  Universality of the oscillations is seen when each $\Delta q$ is scaled appropriately (see text).  Specifically, the plot shows $\Delta q_0 (\gamma/2)$ (blue solid), $\Delta q_2(\gamma)/6$ (red dashed), $-\Delta q_4(\gamma)/8$ (black dotted) and $\Delta q_6(\gamma)/2$ (green dash dotted).}
\label{fig:oscillations}
\end{figure}

We now turn our attention to the oscillations, seen in the numerical evaluation of the integral \eqref{eq:problem:micro:Kmicro} which clearly go beyond the approximations used above.  We isolate the oscillations by looking at the difference between the full calculation and the approximation above (or in other words, the phenomenological result),
\begin{equation}
\Delta q_i=q_{\text{micro},i}-q_{\text{phen},i},
\end{equation}
where $i=0,2,4,6$.  The oscillations (as a function of $\Delta k\,l_{ee}$) in the final result come from integrals of the cosine terms within the factor $\Gamma_{WL}$ given in Eq.~\eqref{eq:gammaWL}.  Noting that there is a certain symmetry in $\Gamma_{WL}$ between interchange of $x_a$, $x_b$ and $x_c$, and singling out the simplest oscillating term for each power of $b$, we find that we would expect the dominant contributions to come from the term with $\cos(2\Delta k x_a)$ in $q_0$; $6 \cos(\Delta k x_a)$ in $q_2$; $-8 \cos(\Delta k x_a)$ in $q_4$; and $2\cos(\Delta k (x_c-x_b))=2\cos(\Delta k x_a)$ in $q_6$.  Hence by scaling each $\Delta q_i$ by an appropriate factor, we expect to see that the oscillations in each $q$ function are identical -- which is demonstrated in Fig.~\ref{fig:oscillations}.  The slight differences follow from the sub-leading terms in $\Gamma_{WL}$.

\begin{figure}
\begin{center}
\includegraphics[width=0.38\textwidth]{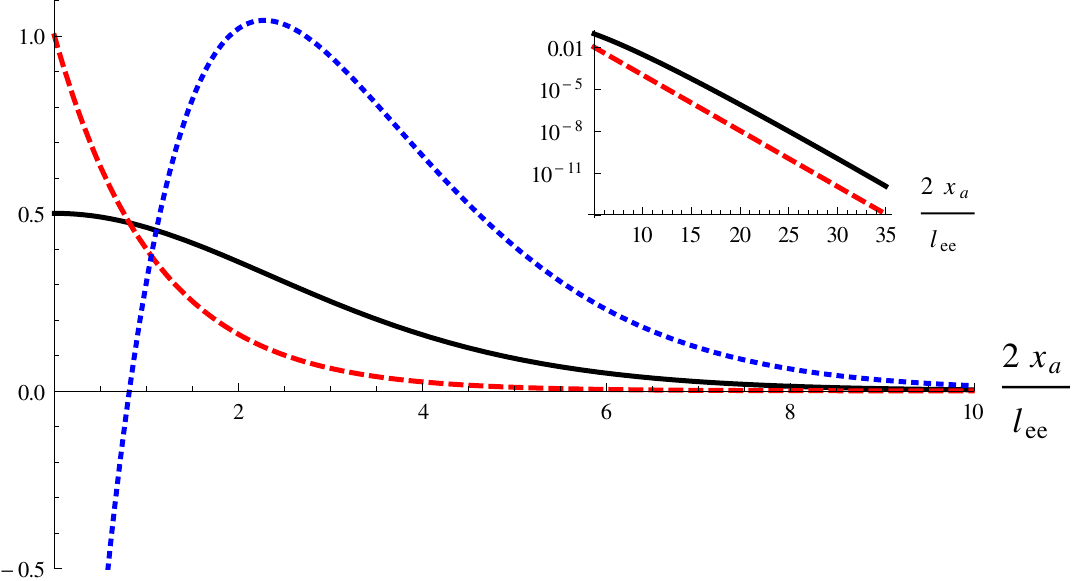}
\end{center}
\caption{[Color online] Comparison between the phenomenological product of Green function, $\mathcal{F}_\text{phen}(x_a)$ (red dashed curve) and the true microscopic curve $\mathcal{F}_\text{micro}(x_a)$ (black solid curve), as defined by Eq.~\eqref{eq:feature_functions}  The blue dotted curve shows the difference between them (multiplied by a factor of $5$ for clarity).  We see that as compared to the phenomenological dephasing model, the true microscopic model has enhanced spectral weight for distances of the order of a few times the dephasing length.  The inset shows the comparison between the same microscopic and phenomenological functions on a longer length scale and a logarithmic axis. }
\label{fig:feature}
\end{figure}

From this demonstration of the universality of the weak localization oscillations, we see that to capture their essence, we can suffice by approximating
\begin{equation}
\Gamma_{WL}(\Delta k x_a,\Delta k x_b,\Delta k (x_a+x_b)) \sim \cos(\Delta k x_a).
\end{equation}
It is then clear why the phenomenological approximation does not produce any weak localization oscillations, as
\begin{equation}
\int_{0}^{\infty}\frac{dx_{a}}{l} e^{ - \frac{2 x_a}{l_{ee}}} \cos(\Delta k\,x_a) = \frac{l_{ee}}{2l} \frac{1}{1+(\Delta k\,l_{ee}/2)^2}
\end{equation}
and the only role of the band-splitting is to renormalize the effective dephasing time.  In fact, carefully collecting all such factors explains immediately the form of the phenomenological result, Eq.~\eqref{eq:problem:phen:Kphen}.  Another important point can be drawn from the above formula however -- which is that such contributions to the weak localization correction mathematically take the form of the Fourier transform of the single particle Green function (actually, a product of six such Green functions, but this distinction is unimportant in the following qualitative analysis).  To have oscillations as a function of $\Delta k$ in the final result, the Green function should therefore have a peak at some finite value of $x_a/l_{ee}$.  In order to see how this comes about, we consider the microscopic kernel Eq.~\eqref{eq:problem:micro:Kmicro} as a function of $x$ at fixed $y=0$, and compare this with the approximation above that yields the exponential -- more precisely, we consider the following two functions:
\begin{eqnarray}\label{eq:feature_functions}
\mathcal{F}_\text{micro}(x_a)&=&\frac{\pi}{4}e^{2x_a/l_{ee}}\int_{-\infty}^\infty dz \frac{\mathcal{R}(e^{2x_a/l_{ee}}-1,z)^2 \mathcal{R}(0,z)}{\cosh^2(\pi z)},  \nonumber \\
\mathcal{F}_\text{phen}(x_a)&=& e^{-2x_a/l_\phi},
\end{eqnarray}
where $\mathcal{R}(x,z)$ is given by Eq.~\eqref{eq:defR}.  The value of $l_\phi$ used in $\mathcal{F}_\text{phen}$ is chosen to best match the exponentially decaying tail of $\mathcal{F}_\text{micro}$ at intermediate length scales (see the inset to Fig.~\ref{fig:feature}), and  matches the $l_\phi$ as fitted in the main text.  We emphasize that this is however a phenomenological fit to intermediate length scales which dominate the integrals for the weak localization correction; and not the true asymptotic of the microscopic function, which involves power law prefactors.

We now turn to shorter length scales, of the order of several $l_{ee}$; in this range, the two functions are plotted in Fig.~\ref{fig:feature}, along with a magnified view of the difference between them.  We see that as compared to the phenomenological model, the true microscopic model has correlations which give an enhanced spectral weight at distances of order a few times the dephasing length.  Having identified mathematically this feature that is responsible for the weak localization oscillations, we can finally ask the question: which property of the correlations in the interacting two leg ladder are responsible for this enhanced spectral weight?

\begin{figure}
\begin{center}
\includegraphics[width=0.35\textwidth]{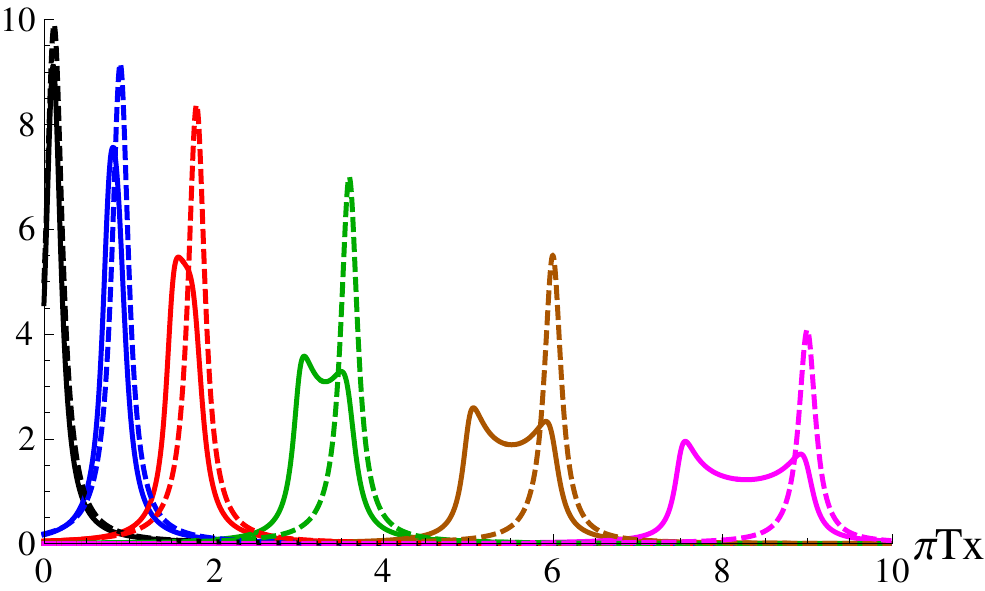}
\end{center}
\caption{[Color online] Real time propagation of spectral function $|\Im[\mathcal{G}_\text{phen/micro}(x,t)]|$ for interaction strength $\alpha=0.2$ and times $\pi T t=0.1,0.75,1.5,3,5,7.5$ (delta functions are broadened to show the correct spectral weight under them). The microscopic model (solid lines) propagates with two peaks due to the two velocities, but the spectral function remains large between these peaks, explaining the extra spectral weight at intermediate distances unaccounted for by the phenomenological model (dashed lines).}
\label{fig:spectralfn}
\end{figure}

To answer this question, we consider the real time propagation of the true microscopic Green function, which is given by an analytic continuation of Eq.~\eqref{eq:GFs1} and yields
\begin{multline}
\mathcal{G}_\text{micro}(x,t)  \\  \propto \left( \frac{(\pi T)^2}{(1+\alpha) \sinh\pi T(x-t) \sinh \pi T(\frac{x}{1+\alpha}-t)}\right)^{1/2};
\end{multline}
and compare it to the real time propagation of the phenomenological Green function, given by a (finite temperature) Fourier transform of Eq.~\eqref{eq:problem:phen:GF-dephasing},
\begin{equation}
\mathcal{G}_\text{phen}(x,t) \propto - \frac{\pi T \, e^{-\pi T \alpha x/2}}{\sinh \pi T(\frac{x}{1+\alpha}-t)}.
\end{equation}
In both cases, we have set $v_F=1$, and have made use that both the inelastic scattering length and the difference in pseudo-spin and charge velocities are proportional to the interaction strength $\alpha$.

The time evolution of these functions is shown in Fig.~\ref{fig:spectralfn}.  In the microscopic model, it is easy to see the propagation via two wavefronts, reflecting the different charge and pseudo-spin velocities; however it also has the curious feature that there is significant spectral weight between the two peaks which is not accounted for in the phenomenological picture.  It is exactly this feature which after integrating over time gives enhanced spectral weight at intermediate distances, and is therefore responsible for the oscillations in the weak localization.  At longer distances $x_a \gg l_{ee}$, this physical property is also responsible for the power law enhancement of $\mathcal{F}_\text{micro}$ as compared to  $\mathcal{F}_\text{phen}$ -- but as these distances contribute an exponentially small amount to the weak localization, we do not comment any more upon this asymptotic behavior.


\end{document}